\documentclass[prb,reprint,showpacs,amsmath,amssymb,superscriptaddress]{revtex4-1}
\usepackage{amsfonts}
\usepackage{graphicx}
\usepackage{amsmath}
\usepackage{float} 
\usepackage{amssymb}
\usepackage{color}
 \usepackage[version=3]{mhchem}
\usepackage[colorlinks,bookmarks=false,citecolor=blue,linkcolor=red,urlcolor=blue]{hyperref}

\begin{document}
\title{Phase Diagrams of Antiferromagnetic Spin-1 Bosons on Square Optical Lattice with  the Quadratic Zeeman Effect}

\author{L. de Forges de Parny}
\email[Corresponding author: ]{laurent.de.forges@physik.uni-freiburg.de}
\affiliation{Physikalisches Institut, Albert-Ludwigs Universit\"{a}t
  Freiburg, Hermann-Herder-Stra{\ss}e 3, D-79104, Freiburg, Germany}
\author{V.$\,$G.  Rousseau}
\affiliation{Physics Department, Loyola University New Orleans, 6363
  Saint Charles Ave., LA 70118, USA}

\begin{abstract}

We study the quadratic Zeeman effect (QZE) in a system of antiferromagnetic spin-1 bosons  on a square  lattice and derive the 
ground-state phase diagrams by means of quantum Monte Carlo simulations and mean field treatment.
The QZE imbalances the populations of the magnetic  sublevels $\sigma=\pm1$ and $\sigma=0$, and therefore affects the magnetic and mobility properties of the 
phases. Both methods show that the tip of the even Mott lobes, stabilized by singlet state, is destroyed when turning on the QZE,
thus leaving the space to the superfluid phase. Contrariwise, the tips of odd Mott lobes remain unaffected.
Therefore, the Mott-superfluid transition with even filling strongly depends on the strength of the QZE, 
and we show that the QZE can act as a control parameter for this transition at fixed hopping.
Using quantum Monte Carlo simulations, we elucidate the nature 
of the phase transitions and examine in detail the nematic order:
the first-order Mott-superfluid transition with even filling  observed in the absence of QZE becomes second order for weak QZE, in contradistinction 
to our mean field results which predict a first-order transition in a larger range of QZE. 
Furthermore, a spin nematic order with director along the $z$ axis is found in the odd Mott lobes and in the superfluid phase for energetically favored $\sigma=\pm1$ states.
In the superfluid phase with even filling, the $xy$ components of the nematic director remain finite only for moderate QZE.

\end{abstract}


\maketitle

\section{Introduction}
\label{sec_1}

The use of ultracold  bosonic systems as  simulators  
for  the Bose-Hubbard model,  proposed in 1998,\cite{Jaksch_1998}
 quickly led to an experimental  realization of the Mott-superfluid transition in 2002.\cite{Greiner02}
For the first time, ultracold gases experiments allowed the simulation  of  well-known
condensed-matter systems\cite{Fisher_1989} and strongly correlated states -- e.g., Mott insulating  state --
thus connecting two so far distinct fields: atomic physics and condensed-matter physics.
Since then,  ultracold atoms in optical lattices are referred to as  quantum simulators, 
i.e., highly tunable systems suitable for the exploration of quantum statistical
lattice models,\cite{bloch08} such as the Bose \cite{Greiner02} and
Fermi \cite{Chin_2006, Schneider08,Jordens08}  Hubbard models.

Since the seminal work of Jaksch \textit{et al.} in 1998,\cite{Jaksch_1998} the motivation for considering 
many species of atoms, or internal degree of freedom, has emerged from the promising perspectives of 
observing coexisting phases, quantum magnetism,  and spin dynamics.
\cite{Dalibard_Gerbier_2011, Gerbier_2006_qnegatif, Stamper_Kurn_2013, Zhao_2015, Lewenstein_Sanpera_1998, Kawaguchi_Ueda_2012, Krutitsky_2016, Kuklov_Svistunov_2003, Chang_2005, Lamacraft_2010}
Indeed, mutlicomponent Bose-Einstein condensates, also called spinor condensates, are ideal systems for engineering quantum phase transitions,\cite{Liu_2016} 
entanglement, metrology,\cite{Esteve_2008} thermometry,\cite{Weld_Lingua}
 and  have possible applications in astrophysics\cite{Alpar_1984_Lattimer_2004} and in quantum chromodynamics.\cite{Son_2001_2002_Tylutki_2016}
The  spin degrees of freedom allow the study of multi band condensed
matter Hamiltonians and the interplay between magnetism and
superfluidity.\cite{Vengalattore08,Vengalattore10}
Contrary to  the standard Bose-Hubbard model with U(1) symmetry, spinor condensates
in optical lattice are described by  the extended Bose-Hubbard model where  spin-spin
interactions introduce an additional symmetry.\cite{Ho_1998,ohmi98} 
The spontaneous breaking of this additional symmetry leads to the establishment of quantum
magnetism in the Mott insulating and superfluid phases, and to multiple transitions.
\cite{Imambekov_2004, snoek04, Capogrosso_Sansone_2010, Deforges_2013, deforges2014, deforges2015, deforges2016}
The on-site spin-spin interaction can be tuned by
using  Feshbach resonance \cite{Theis} and 
the nature of the
spin-spin interaction can be either ferromagnetic (e.g., $^{87}{\rm Rb}$) or
antiferromagnetic (e.g., $^{23}{\rm Na}$) depending on the relative
magnitudes of the scattering lengths in the singlet and quintuplet
channels.\cite{Stamperbook}

The ground-state phase diagram has been intensively investigated using many methods: static and dynamical mean-field 
theory,\cite{pai08, tsuchiya05, kruti04, Li_2016} variational Monte Carlo,\cite{toga} 
analytical,\cite{Imambekov_2004, demler02, Katsura} strong coupling expansion, \cite{Kimura13} density matrix renormalization group, 
\cite{Rizzi05_dmrg, Bergkvist06_dmrg} and quantum Monte Carlo simulations.\cite{apaja06, batrouni2009, Deforges_2013}
Similar to the standard Bose-Hubbard model, the system adopts Mott-insulating
(MI) phases, when the filling is commensurate with the lattice size
and for large enough repulsion between particles, and a superfluid
phase (SF) otherwise.  
The richness of these systems comes from the
magnetic behavior of these phases. 
The two-dimensional case with antiferromagnetic spin-spin interactions is particularly interesting:
a singlet state, with vanishing  local magnetic moment, is observed in the MI phases with even filling
and a nematic state, i.e. a non trivial state breaking spin-rotation symmetry without magnetic order, is observed otherwise.
These magnetic properties are well described by the  bilinear-biquadratic Heisenberg model
in the strong-coupling limit  at integer filling.\cite{Imambekov_2004,Kawashima02,Tsuchiya04, DeChiara_2011, Zhou_2004}
In the weak-coupling limit, the magnetic properties of the superfluidity can be investigated  
within the single mode approximation, where  spin and spatial degree of freedom are decoupled.\cite{Jacob_2012, Zibold_2016, Black_2007, Liu_PRL_2009}
Moreover, the phase transitions are affected by the spin-spin interactions:
the MI-SF transition is first order for even densities (second order otherwise),\cite{Liu_2016, kruti05, pai08, tsuchiya05, toga, Deforges_2013} 
and a singlet-nematic transition occurs inside the MI phases with even densities for 
small spin-spin interactions.\cite{Imambekov_2004, snoek04, Deforges_2013, deforges2014}

The addition of an  external magnetic field completely changes the picture:
the square of the external magnetic field $B^2$, coupled with the square of the local magnetic moment $S_z^2$, 
lifts the degeneracy between the magnetic sublevels $\sigma=\pm1$ and $\sigma=0$, and thus 
constraints the populations of these sublevels. 
This effect, called  the  \textit{quadratic} Zeeman effect in the literature, does not lift the degeneracy 
between  the states $\sigma=+1$ and  $\sigma=-1$, contrary to the Zeeman effect. 
Therefore, the square of the magnetic field in the QZE is equivalent to the impurities in the Blume-Capel model in condensed-matter physics.\cite{Blume_Capel}
This QZE, which has been mostly studied in dilute gas within the single mode approximation, 
acts as a control parameter for the magnetic structure of the superfluidity.\cite{Jacob_2012, Zibold_2016, Jiang_2014, Zhao_2015}
Our study completes the picture for strong interacting particles in optical  lattices.
As shown below, the QZE not only affects the spin degrees of freedom, but
also impacts the mobility of the particles and, therefore, the phase coherence.  
Furthermore, it was experimentally observed that
the nature of the MI-SF transition with even densities is strongly affected by the QZE.\cite{Liu_2016}

The purpose of this paper is to extend our preliminary study\cite{Deforges_2013} by considering  
the QZE in the antiferromagnetic spin-1
system at zero temperature.
We derive the  phase diagrams and investigate the  magnetic structure, 
thanks to the  correlations functions accessible with the QMC method.
The paper is organized as follows: In Sec.~\ref{sec_2}, we introduce the model
and the methods used to study it.  
The mean-field and QMC phase diagrams are discussed in Sec.~\ref{sec_3} and Sec.~\ref{sec_4}, respectively.
The results obtained by both methods are compared in Sec.~\ref{sec_4}.
In Sec.~\ref{sec_5}, we  summarize these results and give some final remarks.

\section{Hamiltonian and Methods}
\label{sec_2}

\subsection{Spin-1 Bose-Hubbard model with the quadratic Zeeman energy}
\label{sec2_subA}

We consider a system of bosonic atoms in the
hyperfine state $F=1$ characterized by the magnetic quantum number
$F_z=\{\pm1, 0\}$. When these atoms are loaded in an optical lattice,
the system is governed by the extended Bose-Hubbard Hamiltonian:\cite{Mahmud_2013, Imambekov_2004, Ho_1998}
\begin{eqnarray}
  \nonumber
  \mathcal {  \hat H} &=&-t\sum_{\sigma,\langle \bf r,\bf r' \rangle} \left (a^\dagger_{\sigma \bf r}
    a^{\phantom\dagger}_{\sigma \bf r'} + {\rm h.c.}\right ) + \frac{U_0}{2}\sum_{\bf r}
  {\hat n}_{ \bf r} \left ( {\hat n}_{\bf r}-1\right )\\
  &&+ \frac{U_2}{2}\sum_{\bf r} \left({\bf {\hat S}}_{\bf r}^2-2 {\hat n}_{\bf r} \right ) - q   \sum_{\bf r} {\hat n}_{0 \bf r}  ,
\label{Hamiltonian_compact_form}
\end{eqnarray}
where operator $a^{\phantom\dagger}_{\sigma {\bf r}}$ ($a^ \dagger_{\sigma {\bf r}}$)
annihilates (creates) a boson in the Zeeman state $\sigma=\{\pm1, 0\}$ (or $\sigma=\{\downarrow, 0, \uparrow\}$)
 on site ${\bf r}$ of a periodic square lattice of
size $L\times L$.

The first term in the Hamiltonian is the kinetic term which allows
particles to hop between neighboring sites $\langle {\bf r,r'} \rangle$ with strength $t$.
The number operator
$\hat{n}_{ {\bf r}} \equiv \sum_{\sigma} {\hat n}_{ \sigma \bf r} =
\sum_{\sigma} a^\dagger_{\sigma {\bf r}} a^{\phantom{\dagger}}_{\sigma
  {\bf r}}$ counts the total number of bosons on site $\bf r$.
$N_\sigma \equiv \sum_{\bf r} \langle  {\hat  n}_{\sigma {\bf r}} \rangle$ will denote the total
number of $\sigma$ bosons, $\rho_\sigma \equiv N_\sigma/L^2$ the
corresponding density, and $\rho \equiv \sum_\sigma \rho_\sigma$ the total density.  The operator
${\bf {\hat S}}_{\bf r} = ({\hat S}_{x, {\bf r}}, {\hat S}_{y,{\bf r}}, {\hat  S}_{z,{\bf r}})$ is
the spin operator where $ {\hat  S}_{\alpha,{\bf r}}  =\sum_{\sigma,\sigma' }
a^\dagger_{\sigma \bf r} J_{\alpha,\sigma \sigma'} a_{\sigma' \bf r}$,
$\alpha =\{ x,y,z\}$ and the $J_{\alpha, \sigma \sigma'}$ are standard
spin-1 matrices;  hence 
\begin{equation}
\left\{
\begin{array}{lll}
 \vspace*{0.2 cm}
{\hat  S}_{x, {\bf r}}  &=&\frac{1}{\sqrt{2}} \left ( ({a}^\dagger_{\downarrow {\bf r}}+{a}^\dagger_{\uparrow {\bf r}}){a}_{0{\bf r}}^{\phantom\dagger} +  
  {a }^\dagger_{0{\bf r}}({a}_{\downarrow {\bf r}}^{\phantom\dagger}+{a}_{\uparrow {\bf r}}^{\phantom\dagger}) \right ) \label{operateurs_S}~,\\
\vspace*{0.2 cm}
{\hat  S}_{y, {\bf r}} &=&\frac{\textrm{1}}{\sqrt{2}} \left ( ({a }^\dagger_{\downarrow {\bf r}}-{a}^\dagger_{\uparrow {\bf r}}){a}_{0{\bf r}}^{\phantom\dagger} -  
 {a}^\dagger_{0{\bf r}}({a}_{\downarrow {\bf r}}^{\phantom\dagger}-{a}_{\uparrow {\bf r}}^{\phantom\dagger}) \right ) ~,\\
{\hat S}_{z, {\bf r}} &=&{\hat n}_{\uparrow {\bf r}} -{\hat n}_{\downarrow {\bf r}}~.
 \end{array}
\right.
\end{equation}

The parameters $U_0$ and $U_2$ are the on-site spin-independent and
spin-dependent interaction terms.  The nature of the bosons determines
the sign of $U_2$ and consequently the nature of the on-site spin-spin
interaction, whereas the spin-independent part is always repulsive,
$U_0 > 0$.  The $U_2<0$ case (\rm{e.g.}, $^{87}{\rm Rb}$\cite{Kempen_2002})  maximizes
the  local magnetic  moment 
\begin{eqnarray}
S^2(0)  \equiv \frac{1}{L^2} \sum_{\bf r} \langle   {\hat  {\bf S}}_{{\bf r}}^2 \rangle.
\label{magnetic_local_moment}
\end{eqnarray}
Therefore, this is referred to as the
``ferromagnetic" case, whereas the $U_2>0$ case (\rm{e.g.} $^{23}{\rm Na}$\cite{Burke_1998, Imambekov_2004}), 
which favors minimum local magnetic moment, is referred to as
``antiferromagnetic".
In this article, the hopping parameter sets the energy scale $t=1$ and we focus on 
$^{23}{\rm Na}$ atoms, i.e., $U_2 = 0.036 U_0$.\cite{Burke_1998, Imambekov_2004}

The fourth term in Eq.~(\ref{Hamiltonian_compact_form}) corresponds to the Zeeman energy shift between
 the sublevels $\sigma =\pm1$ and $\sigma =0$ with amplitude $q$.
The parameter $q$ arises either from an applied magnetic field $B$, 
which leads to the quadratic Zeeman energy with $q\sim B^2$, or 
from a microwave field, which in this case leads to a positive or negative $q$ value.\cite{Jiang_2014, Gerbier_2006_qnegatif} 

In the absence of Zeeman shift, i.e., $q=0$, the Hamiltonian $\mathcal {  \hat H}$  Eq.~(\ref{Hamiltonian_compact_form}) 
has the symmetry U(1)$\times$SU(2), associated with the total mass conservation  [U(1) symmetry], 
times the SU(2) symmetry of the spin rotation invariance on the Bloch sphere. For $q\neq0$, the SU(2) 
symmetry is reduced to U(1), leading to a model with global symmetry U(1)$\times$U(1).

Once developed, the ${\bf {\hat  S}}^2_{\bf r}$ operator exhibits
contact interaction terms and conversion terms 
between the Zeeman states, i.e., $a^\dagger_{0 \bf r}a^\dagger_{0 \bf r}a_{\downarrow \bf r} a_{\uparrow \bf r} + {\rm h.c.}$, 
which destroys  a pair of particles respectively in the Zeeman states $\uparrow$ and $\downarrow$ and creates two bosons in the  state $\sigma=0$, and vice-versa.
Therefore, only the total number of atoms $N_{\rm tot} = \sum_{\bf r, \sigma} \langle  {\hat  n}_{\sigma {\bf r}} \rangle$, associated to the U(1) symmetry, is conserved.

\subsection{Mean-Field Approximation and quantum Monte Carlo Simulations}
\label{sec2_subB}

We use a combined  approach based on mean-field theory, supplemented  with numerically  exact
quantum Monte Carlo (QMC) simulations.

Although the mean field approximation does not give quantitatively accurate values for the phase boundaries, 
it is a practical method which allows for the rapid reconstruction of phase
diagrams.\cite{pai08, Imambekov_2004, Deforges_2011}
We use a mean-field formulation based on a decoupling approximation which decouples the
hopping  term  to  obtain  an  effective  one-site  problem. 
Introducing the superfluid order parameter  $\psi_\sigma \equiv \langle
a^\dagger_{\sigma \bf r}\rangle = \langle a_{\sigma \bf r}\rangle$, we replace the creation and destruction operators on
site $\bf r$ by their mean values $\psi_\sigma$.
Since  we  are interested in equilibrium states of spatially uniform superfluids, the order parameters can be chosen to be real.
 Using this ansatz, the kinetic-energy terms, which are nondiagonal in boson creation
and destruction operators, are decoupled as
\begin{eqnarray}
\nonumber
a^{\dagger}_{\sigma \bf r}   a^{\phantom\dagger}_{\sigma \bf r'}
 \simeq (a^{\dagger}_{\sigma \bf r}  + a^{\phantom\dagger}_{\sigma \bf r'})  \psi_\sigma - \psi_\sigma^2   ~.
\label{mfapprox}
\end{eqnarray}
The Hamiltonian  Eq.~(\ref{Hamiltonian_compact_form}) is rewritten as a sum over local terms 
$\mathcal {  \hat H} =  \sum_{\bf r} \mathcal {  \hat H}^{MF}_{\bf r}$, where
\begin{eqnarray}
\nonumber
\mathcal {  \hat H}^{MF}_{\bf r} &=& -z t \sum_\sigma  \left [  (a^{\dagger}_{\sigma {\bf r}}  + a^{\phantom\dagger}_{\sigma {\bf r}})  \psi_\sigma 
   -\psi_\sigma^2 \right ]+ \frac{U_{0}}{2} {\hat n}_{\bf r} \left ( {\hat n}_{\bf r} -1\right )   \\
&& + \frac{U_2}{2}    \left({\bf {\hat  S}}_{\bf r}^2-2 {\hat n}_{\bf r} \right ) - q   {\hat n}_{0 \bf r}  ,
\label{Hamiltonian_MF}
\end{eqnarray}
where $z=4$ is the number of nearest neighbors in a square lattice.
The mean-field Hamiltonian Eq.~(\ref{Hamiltonian_MF})  can be easily
diagonalized numerically in a finite occupation-number  basis $\{ |n_-, n_0, n_+ \rangle \}$, 
with the truncation $n_{max}=10$, 
and  the lowest eigenenergy is minimized with respect to $\psi_\sigma$. 
This gives the order parameters of the ground state
and its eigenvector $|\Psi_{GS}^{MF}  \rangle$.
At zero temperature, the system is in a Bose-Einstein condensed  phase if at  least  one  of  the  order  parameters is  nonzero
and is, otherwise, in an insulating phase.
The condensate fractions and densities are  respectively defined by
\begin{eqnarray}
C^{MF}_{\sigma} \equiv |\psi_\sigma|^2~, 
\label{MF_condensate_fraction}
\end{eqnarray}
\begin{eqnarray}
\rho_{\sigma} &\equiv&\langle \Psi_{GS}^{MF}  | a_\sigma^{\dagger}  a_\sigma |\Psi_{GS}^{MF}  \rangle, 
\label{MF_density}
\end{eqnarray}
with $C^{MF}= \sum_\sigma C^{MF}_{\sigma}$ the total condensate fraction.
\\

Our QMC simulations are based on  the Stochastic Green Function
algorithm\cite{SGF} with directed updates,\cite{directedSGF}
an exact quantum Monte Carlo technique that allows canonical or grand
canonical simulations of the system as well as measurements of
many-particle Green functions.  In particular, this algorithm can
simulate efficiently the conversion terms.

In this work, we used  the canonical formulation where the total number of bosons $N_{\rm tot}$, 
is conserved whereas the individual  number of bosons $N_\sigma$
fluctuate. The QMC algorithm also conserves the total spin along
$z$, $S_{{\rm tot},z}= N_{+} - N_{-}$, which adds a constraint to the
canonical one. The value of $S_{{\rm tot},z}$ in a given canonical simulation
is then fixed by the choice of the initial numbers of particles.
In this paper, we work in the spin sector $S_{{\rm tot},z}=0$.
Due to this constraint, the magnetic physical
quantities, involving $S_{\alpha,{\bf r}}$ operators, are identical
for the $x$ and $y$ axes but may be different for the $z$ axis. 
The initial symmetry, where all three axes should behave identically, is
broken in our simulations.
Using this algorithm we were able to simulate the system
reliably for clusters going up to $L=12$ with $N=288$ particles.
A large enough inverse temperature of  $\beta = 2 L/t$ allows one
 to eliminate thermal effects.

In particular, we calculate  the averaged densities $ \rho_\sigma =  \sum_{\bf r}  \langle {\hat n}_{\sigma {\bf r}} \rangle / L^2$
and the singlet density
\begin{eqnarray}
\label{singlet_density}
\rho_{sg} \equiv  \frac{1}{L^2}  \langle  {\hat A}_{sg}^{\dagger}  {\hat A}_{sg}^{\phantom\dagger}\rangle~,
\end{eqnarray}
where $ {\hat A}_{sg}^{\dagger} = \frac{1}{\sqrt{6}}(2  {\hat a}_{\uparrow }^{\dagger}    {\hat a}_{\downarrow }^{\dagger}   -   {\hat a}_{0 }^{\dagger}    {\hat a}_{0 }^{\dagger} )$, 
which  measures the number of pairs of bosons with  a vanishing magnetic local moment. 
The chemical potential is defined as the discrete
difference of the energy
\begin{equation}
\mu(N_{\rm tot})=E(N_{\rm tot}+1)-E(N_{\rm tot}), \label{mu_deltaE}
\end{equation}
 which is valid in the ground state where the free energy is equal
to the internal energy  $E= \langle  \mathcal {  \hat H}  \rangle $.

The analysis of the magnetic structure
requires the calculation of the  local magnetic  moment 
$S^2(0)  = S_x^2(0)+ S_y^2(0)+ S_z^2(0) $, where
\begin{equation}
S_\alpha^2(0) \equiv  \frac{1}{L^2}\sum_{\bf r}\langle { {\hat  S}}_{\alpha, {\bf r}}^2 \rangle ~,
\label{magnetic_local_moment}
\end{equation}
with components
 \begin{equation}
 \left\{
 \begin{array}{lll}
  \vspace*{0.2 cm}
  {\hat  S}^2_{x, y, {\bf r}}  &=& 
    a_{0 {\bf r}}^\dagger      a_{0 {\bf r}}^\dagger       a_{\uparrow {\bf r}}^{\phantom\dagger}        a_{\downarrow {\bf r}}^{\phantom\dagger}
 +  a_{\downarrow {\bf r}}^\dagger       a_{\uparrow {\bf r}}^\dagger       a_{0 {\bf r}}^{\phantom\dagger}        a_{0 {\bf r}}^{\phantom\dagger}\\
 &+& ({\hat n}_{0 {\bf r}} +\frac{1}{2})({\hat n}_{\downarrow {\bf r}}+{\hat n}_{\uparrow {\bf r}})
 +{\hat n}_{0 {\bf r}}  \pm \mathcal O 
  \vspace*{0.2 cm} \\
    \vspace*{0.2 cm}
 {\hat  S}^2_{z, {\bf {\bf r}}} &=& 
 {\hat n}_{\uparrow {\bf r}}^2+{\hat n}_{\downarrow {\bf r}}^2 -2 {\hat n}_{\downarrow {\bf r}} {\hat n}_{\uparrow {\bf r}}
 \label{magnetic_local_moment_squared}
 \end{array}
 \right.
 \end{equation}
and $\langle \mathcal O  \rangle=0$, $S_x^2(0)=S_y^2(0)$ using our algorithm.
Since local quantities are insufficient for indicating a long range 
magnetic order, one needs to calculate 
the equal-time spin-spin correlation functions
\begin{eqnarray}
  S_{\alpha \alpha}( {\bf R}) \equiv \frac{1}{L^2} \sum_{\bf r} \langle
  {\hat S}_{ \alpha, {\bf r}} {\hat  S}_{ \alpha, {\bf r+R}}  \rangle~, 
   \label{Saa}  
\end{eqnarray}
and the magnetic structure factor
\begin{eqnarray}
\mathcal S_{\alpha \alpha}({\bf k}) \equiv \frac{1}{L^2}\sum_{\bf R} e^{i{\bf k\cdot R}} S_{\alpha \alpha}({\bf R}) ~,
 \label{magnetic_structure_factor}  
\end{eqnarray}
where ${\bf k} = (k_x,k_y)$ and $k_{x,y}$ are integer multiples of $2 \pi /
L$.
The total global magnetization is given by $M^2_{\rm tot} = 2 M_x^2+M_z^2$ 
with $ M_\alpha^2 =\mathcal S_{\alpha \alpha}({\bf k=0})$.
Additionally, we calculate the order parameter of the 
nematic phase associated to a director along the $z$ axis defined by
\begin{eqnarray}
\Theta_{zz} \equiv \frac{1}{L^4} \sum_{\bf r, R} \langle
  {\hat S}^2_{ z, {\bf r}} {\hat  S}^2_{ z, {\bf r+R}}  \rangle~.  
  \label{Szz_square}  
\end{eqnarray}
The quantity  $\Theta_{zz} $ is maximized when the spins align or (randomly) antialign along  the $z$ axis.

We also calculate the one body Green functions
\begin{equation}
  G_\sigma({\bf R}) =\frac{1}{2L^2}\sum_{\bf r} \langle a^\dagger_{\sigma {\bf
      r+R}}a^{\phantom{\dagger}}_{\sigma {\bf r}} + a^\dagger_{\sigma {\bf
      r}}a^{\phantom{\dagger}}_{\sigma {\bf
      r+R}}\rangle~,
\label{green_onebody}
\end{equation}
which measure the phase coherence of particles in Zeeman state $\sigma$.  
The density of $\sigma$ bosons  with zero momentum  -- here after called the condensate -- is defined by 
\begin{equation}
  C_\sigma =\frac{1}{L^2}\sum_{\bf R} G_\sigma({\bf R})~.
\label{condensate_fraction_QMC}
\end{equation}

Finally, the superfluid density
is given by \cite{Ceperley_1989}
\begin{equation}
\rho_s= \frac{\langle W^2 \rangle}{4t\beta},
\label{spinone_rhosc2D}
\end{equation}
where the total winding number $W=W_{-}+W_0+W_{+}$ is a topological
quantity.

\section{Mean-Field Phase Diagrams}
\label{sec_3}

The minimization of the free energy of Hamiltonian Eq.~(\ref{Hamiltonian_compact_form}) leads to a competition between 
the local magnetic moment $S^2(0)$ (dominant at low $|q|$) and the Zeeman term  (dominant at large $|q|$).
At large $|q|$, the minimization of the Zeeman term leads to an occupation of state $\sigma=0$ ($\sigma=\{\downarrow, \uparrow\}$) for positive (negative) $q$. 
As we will show,  this competition leads to interesting effects for filling $\rho=2$ for which the 
local magnetic moment Eq.~(\ref{magnetic_local_moment}) could be fully minimized, i.e., $S^2(0)=0$.

\subsection{Phase Diagrams}
\label{sec3_subA}

\begin{figure}[h]
	\includegraphics[width=1 \columnwidth]{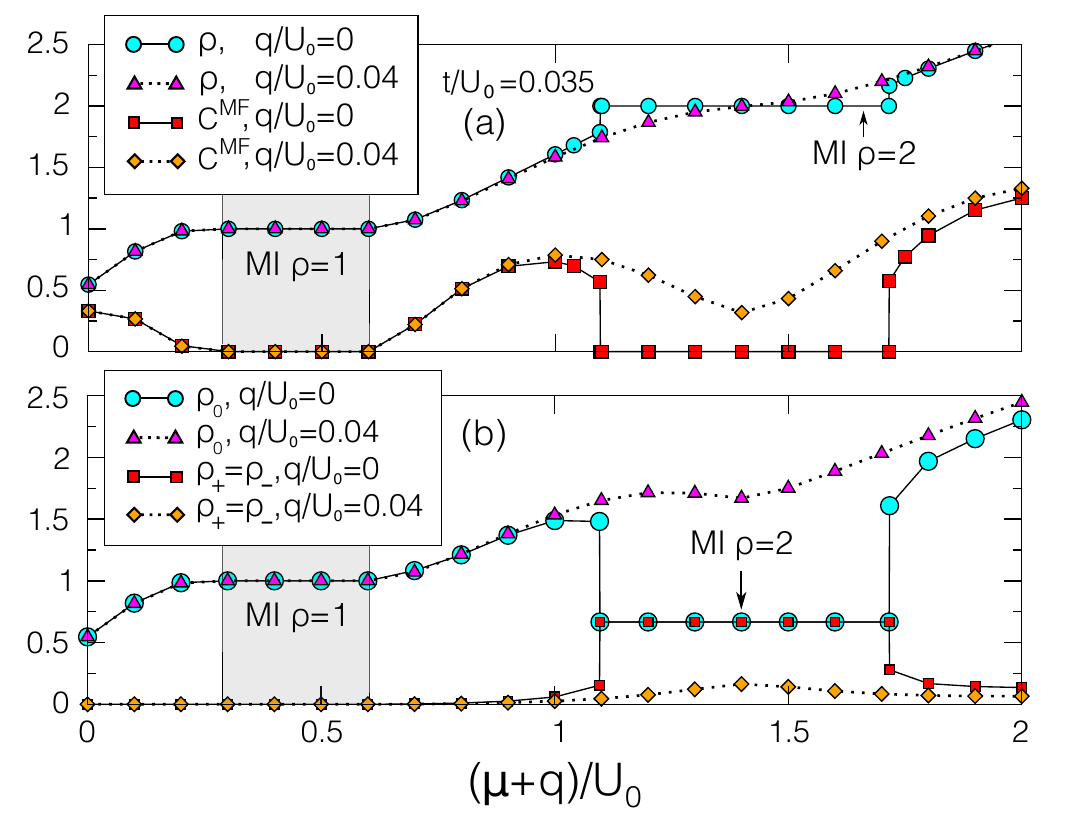}
        \caption {(Color online) Mean field data at fixed $t/U_0=0.035$ for $q=0$ and $q/U_0=0.04$. 
        (a) The Mott phase, indicated by the $\rho=2$ plateau with $C^{MF}=0$ observed for $q=0$,  disappears for $q/U_0=0.04$.
        (b) The density $\rho_0$ increases with $q$.
        The jumps in the densities and in the condensate fraction indicate  first-order MI-SF transitions. 
        }
\label{Figure1}
\end{figure}
In the no-hopping limit, $t/U_0 \to 0$,  the charge gap is easily calculated:
the Mott phase with one boson per site has an energy $\varepsilon(\rho=1)=0$ for $q\leq0$ and $\varepsilon(\rho=1)=-q$ for $q>0$,
whereas the Mott phase with two bosons per site has an energy 
\begin{eqnarray}
\nonumber
\varepsilon(\rho=2) &=& U_0 \left(1-\frac{1}{2}\sqrt{4\left (\frac{q}{U_0} \right )^2-4q \frac{U_2}{U_0^2}+9\left (\frac{U_2}{U_0} \right )^2} \right)  \\ &-& q-\frac{U_2}{2}~,
\end{eqnarray}
associated to the non-degenerate wave function
\begin{eqnarray}
\nonumber
|\Phi_{\rho=2} \rangle &=& \alpha  |1,0,1\rangle \\
&-&\alpha \left(  \frac{\sqrt{4q^2-4qU_2+9U_2^2} +2q-U_2}{2\sqrt{2}U_2}  \right) |0,2,0\rangle, \ \ \ \  \ \ \
\label{Local_wave_function_rho2}
\end{eqnarray}
 obtained by diagonalizing  Eq.~(\ref{Hamiltonian_compact_form}) in the Fock basis 
$|n_-,n_0,n_+\rangle =\{  |1,0,1\rangle, |0,2,0\rangle \}$ with $t=0$, and
$\alpha$ ensuring the normalization of the wave function.
\begin{figure*}[t]
\hbox{\includegraphics[width=1 \columnwidth]{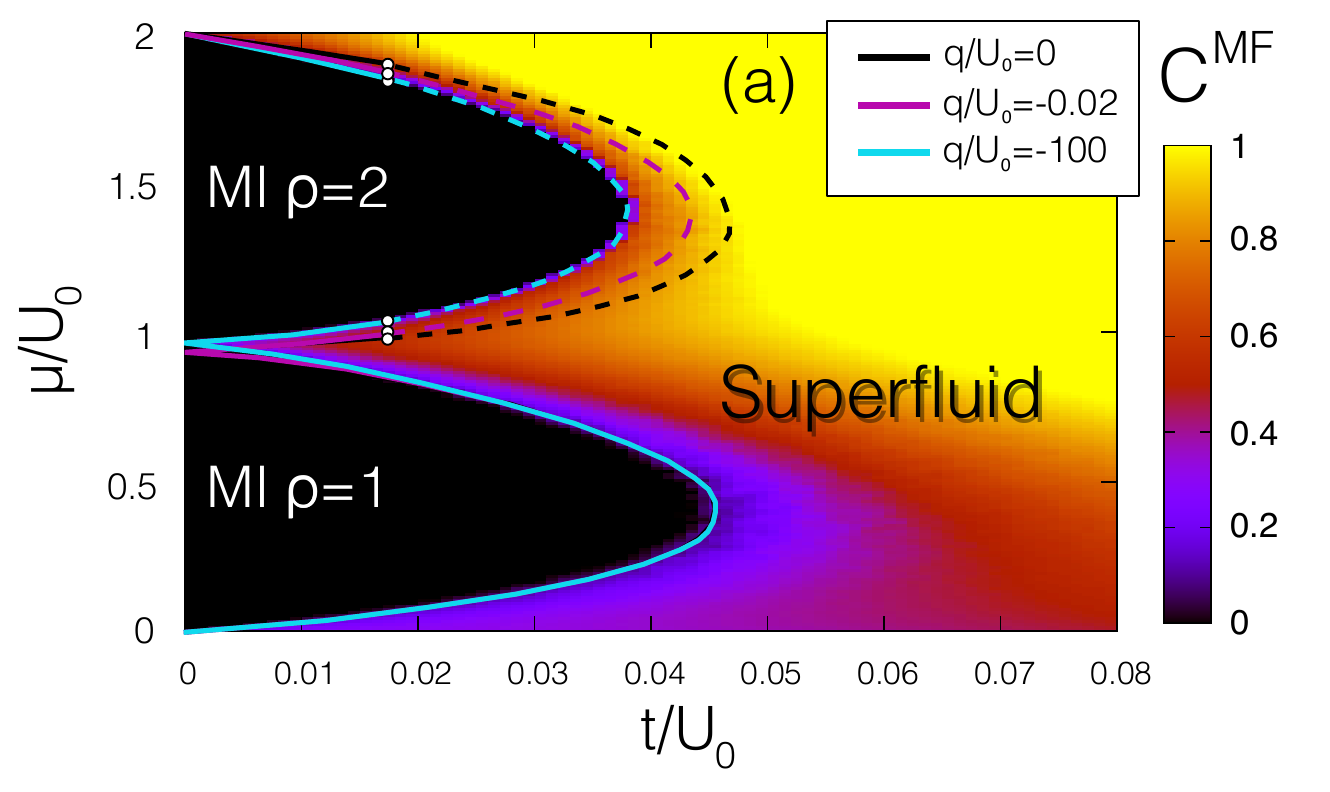}
  \includegraphics[width=1 \columnwidth]{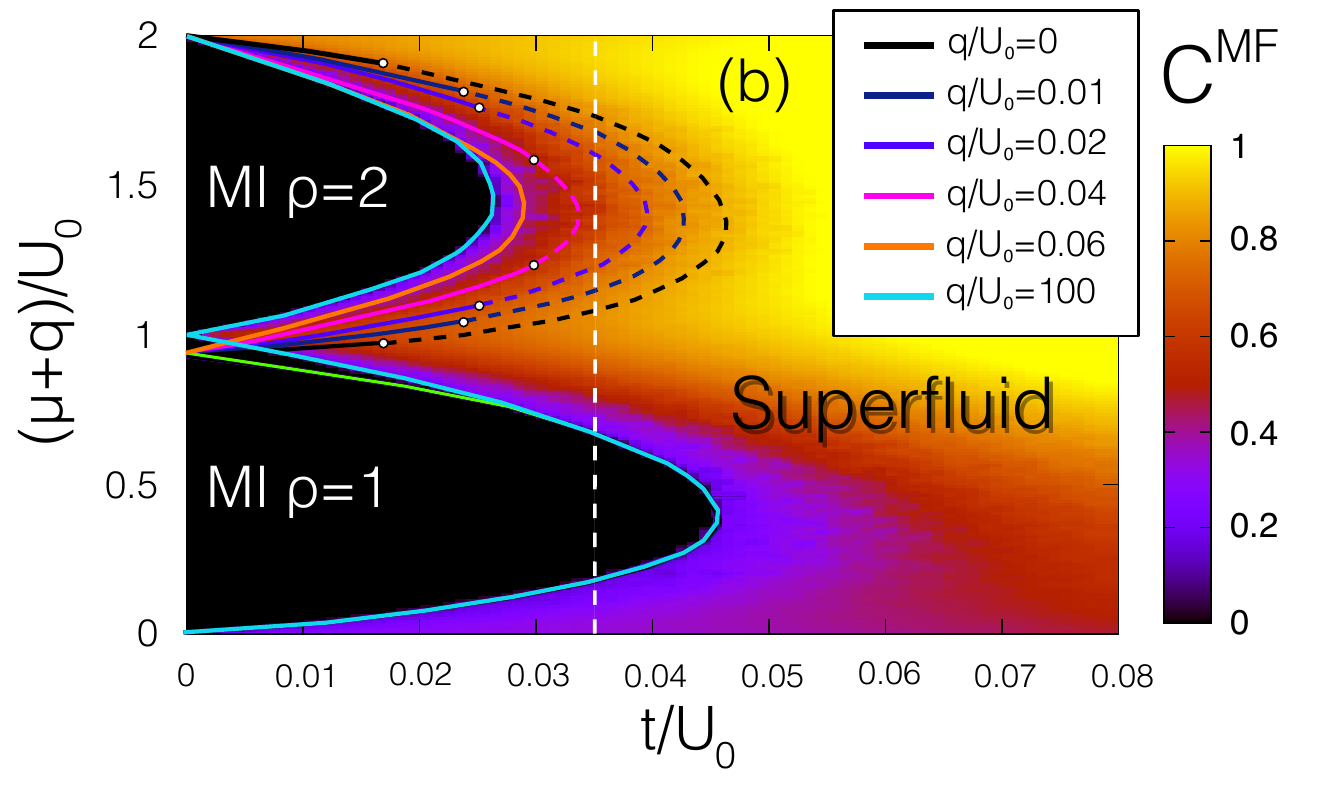}}
        \caption {(Color online)  Mean-field phase diagrams for $U_2/U_0=0.036$ with (a) negative $q$ and (b) positive $q$ values. 
        Contrary to the $\rho=1$ Mott lobe, $q$ strongly affects the tip of the $\rho=2$ Mott lobe.
        The dashed (plain) lines indicate a first- (second-) order transition, and white dots are tricritical points.
        The vertical dashed line in (b) at $t/U_0=0.035$ corresponds to the cut shown in Fig.~\ref{Figure1}.
        False colors show the condensate fraction $C^{MF}$ for (a) $q/U_0=-100$ and (b) $q/U_0=100$. 
        }
        \label{Figure2}     
\end{figure*}
For $q=0$, the on-site energy $\varepsilon(\rho=2, q=0)=U_0-2U_2$ is minimized by minimizing $S^2(0)=0$, that is by
adopting a the singlet state given  by $|\Phi_{\rho=2, q=0}\rangle = \frac{1}{\sqrt{3}} \left(   \sqrt{2} |1,0,1\rangle  -|0,2,0\rangle  \right)$.
In the  limits $q\to \pm\infty$, the wave function reads $ |\Phi_{\rho=2, q\to -\infty}\rangle =  |1,0,1\rangle$ with energy
$\varepsilon(\rho=2) = U_0-U_2$ and $ |\Phi_{\rho=2, q\to +\infty}\rangle =  -|0,2,0\rangle$ with energy
$\varepsilon(\rho=2) = U_0-2q$.
Therefore, for $q=0$,  the base of the Mott lobes for odd
filling is $\Delta \mu/U_0=1-2U_2/U_0$, whereas it is $\Delta
\mu/U_0=1+2U_2/U_0$ for even filling.  The even lobes grow at the
expense of the odd ones, which disappear entirely for $U_2/U_0=0.5$.
In the limit  $q\to - \infty$, the base of the Mott lobes for odd
filling is $\Delta \mu/U_0=1-U_2/U_0$ which disappear entirely for $U_2=U_0$, 
whereas it is $\Delta \mu/U_0=1+U_2/U_0$ for even filling. 
For  $q\to  \infty$, we recover the standard single-species Bose-Hubbard model  
where the  base of the Mott lobes  is $(\Delta \mu +q) /U_0=1$ for all  filling ($q$ is absorbed in the chemical potential).

When turning on the hopping $t\neq0$,  the phase diagram is calculated by plotting the total density $\rho$ and the total condensate fraction $C^{MF}$ versus $\mu/U_0$
for many hopping $t/U_0$ with fixed $q/U_0$ value.
An example of such a vertical slice in the phase diagram is plotted in Fig.~\ref{Figure1}(a) for $t/U_0=0.035$.
We see that all compressible regions, $\kappa \equiv \partial \rho / \partial \mu \neq 0$, are superfluid with $C^{MF}\neq0$ while the incompressible
plateaus, $\kappa= 0$, are not superfluid, they are the Mott insulators. 
Figure~\ref{Figure1}(a) also shows that increasing $q/U_0$ has a strong effect for $\rho>1$: 
the charge gap of the MI with $\rho=2$, clearly observed for $q=0$, vanishes for  $q/U_0=0.04$.
Furthermore, the population $\rho_\pm$ and $\rho_0$ are  sensitive to $q$: 
the population of state $\sigma=0$ ($\sigma=\{\downarrow, \uparrow\}$)  increases (decreases) with $q$, as expected [Fig.~\ref{Figure1}(b)].
The jump in the densities and in the condensate fraction indicate a first-order Mott-superfluid transition.\cite{pai08, Deforges_2013}

 As $t/U_0$ increases, the MI regions are reduced and eventually disappear. 
Outside the MI the system is superfluid. 
The evolution of the mean-field phase diagrams with respect to $q/U_0$ is plotted in Fig.~\ref{Figure2}.

In the $\rho=1$ Mott lobe,  the local magnetic moment is fixed at $S^2(0)=2$ and the densities are 
$\rho_-=1$ or $\rho_+=1$ for $q<0$ and $\rho_0=1$ for $q>0$. Clearly, the tip of the  $\rho=1$ Mott lobe, which ends at $t_c/U_0\simeq 0.043$,
does not vary with $q$.
The situation is very different for $\rho=2$, where the possible minimization of $S^2(0)$ competes with both the 
QZE and the kinetic term. For $q=0$, the tip of the $\rho=2$ Mott lobe is stabilized by the creation of the singlet state with $S^2(0)=0$.
Since the QZE destroys the singlet state, hence $S^2(0)\neq0$, we expect the superfluid region for $q\neq0$ to grow
at the expense of the $\rho=2$ Mott region.
This effect is clearly observed for  both negative  and  positive $q$ values in Fig.~\ref{Figure2}(a) and \ref{Figure2}(b). 
Furthermore, the nature of the  Mott-superfluid transition varies with $q$ for $\rho>1$: 
first- (second-) order transitions are denoted by dashed (plain) lines.

\subsection{MI-SF transition versus QZE}
\label{sec3_subB}

\begin{figure}[h]
	\includegraphics[width=1 \columnwidth]{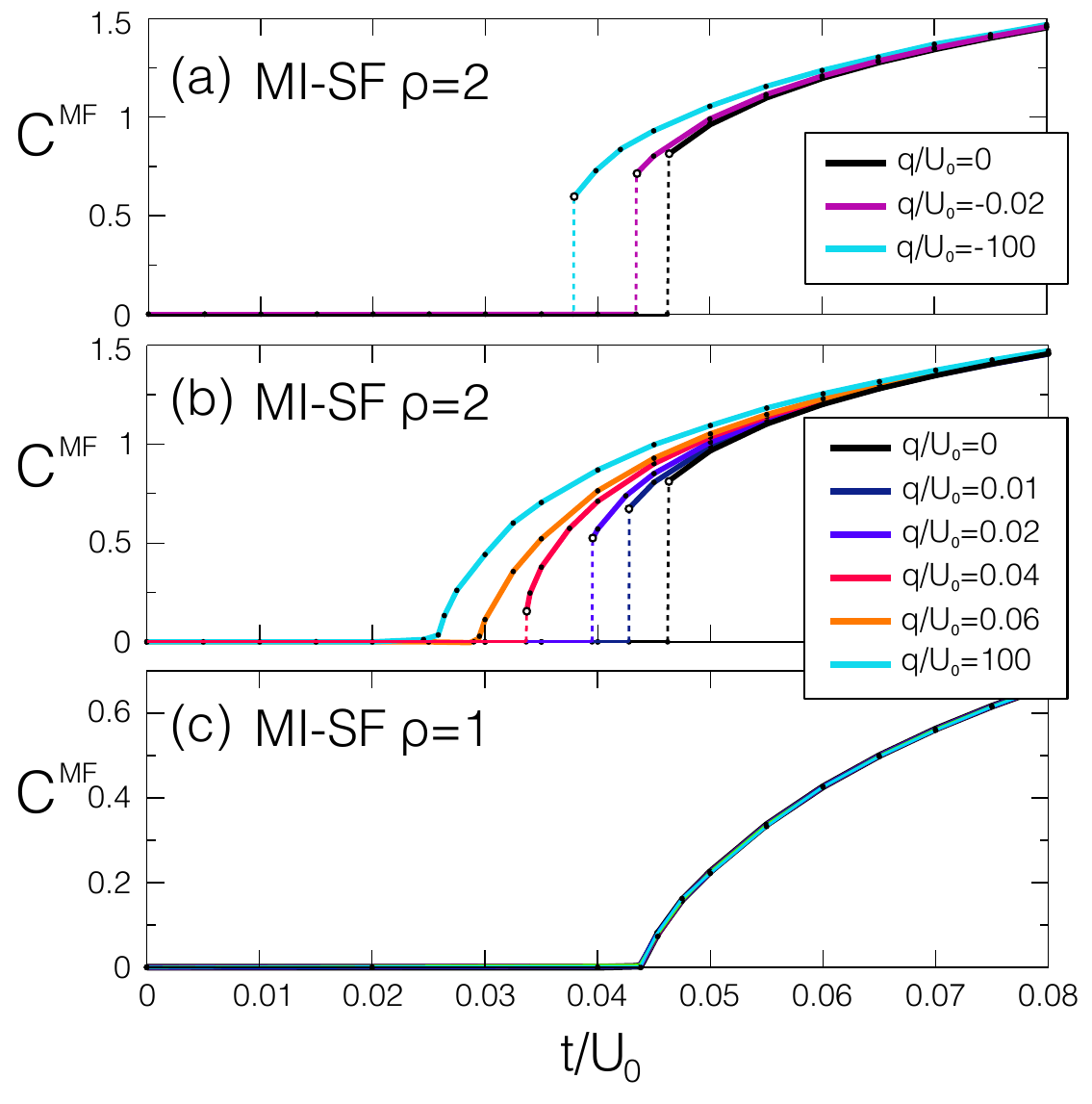}
        \caption {(Color online) Mean-field Mott-superfluid transition for (a), (b) $\rho=2$ and (c) $\rho=1$. 
        The jump in $C^{MF}$ (vertical dashed lines) indicates a first-order transition. 
        The transition is second order otherwise.}
\label{Figure3}
\end{figure}
We first focus on the Mott-superfluid transition at fixed integer filling; see Fig.~\ref{Figure3}.
As discussed before, the cases $\rho=1, 2$ exhibit different behaviors, since 
the singlet state is destroyed by $q$ for  $\rho=2$.
Clearly, $q$ has no quantitative effect on the total condensate fraction $C^{MF}$ for $\rho=1$  since 
the local magnetic moment is fixed to $S^2(0)=2$ and is insensitive to $q$; see Fig.~\ref{Figure3}(c).
Nevertheless, $q$ affects the distribution of the condensate populations: $C_0^{MF}=0$ for $q<0$, whereas
$C_\pm^{MF}=0$ for $q>0$ (not shown).
Similar to the standard single species Bose-Hubbard model using the same mean field formulation,  
the transition takes place at $t_c/U_0\simeq 0.043$.
For $\rho=2$, the destruction of the singlet state for $q\neq0$ leads to a shift of $t_c/U_0$
toward smaller values and the superfluid region grows at the expense of the  Mott phase; see  Figs.~\ref{Figure3}(a) and \ref{Figure3}(b).
We recover the single species Bose-Hubbard model with 
second order phase transition at $t_c/U_0\simeq 0.025$ for $q\to \infty$; see Fig.~\ref{Figure3}(b).
The transition remains first-order for $q<0$, whereas the transition becomes continuous for $q/U_0>0.04$.
Here also, $q$ affects the distribution of the condensate populations for which $C_0^{MF}=0$ for $q<0$, whereas
$C_\pm^{MF}=0$ for $q>0$ (not shown).
\begin{figure}[t]
	\includegraphics[width=1 \columnwidth]{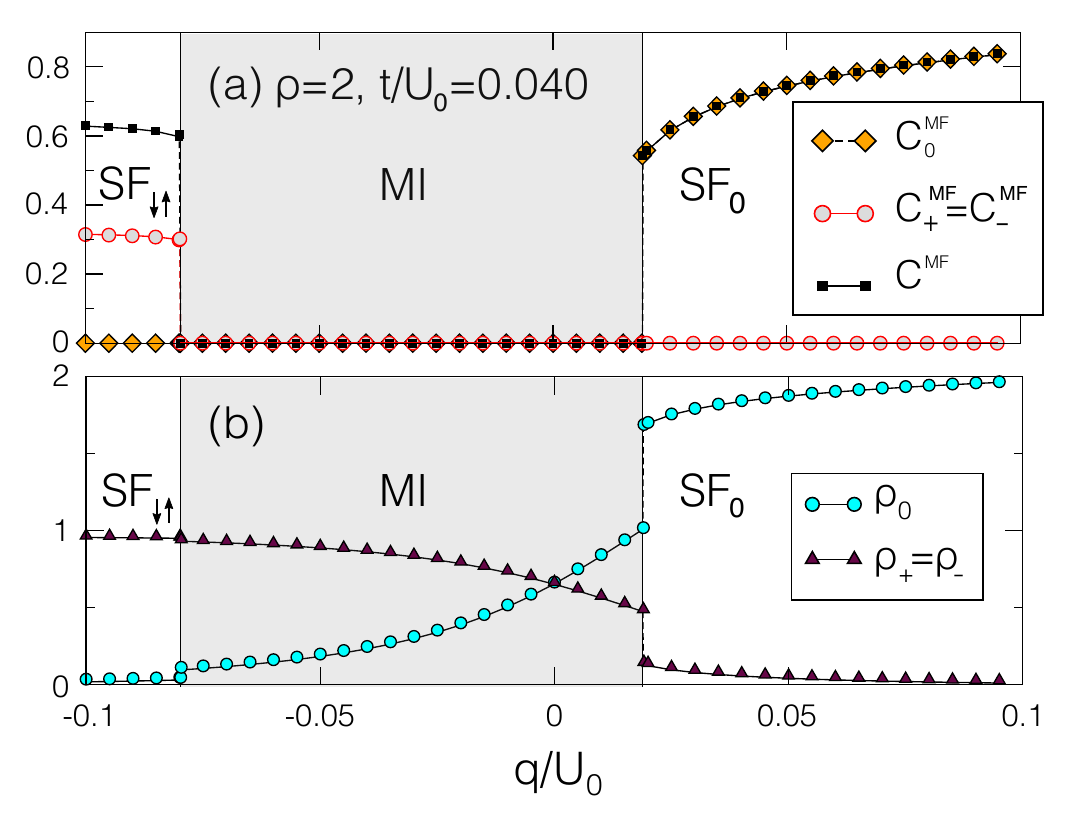}
        \caption {(Color online) Mean-field data: $q$ acts as a control parameter for the $\rho=2$ MI-SF transition at fixed hopping $t/U_0=0.04$.
        For $q\to -\infty$ the system is in the SF$_{\downarrow \uparrow}$ with $C_0^{MF}=0$ and enters in the MI phase at $q/U_0=-0.08$ when increasing
        $q$. Then, the system suddenly adopts a SF$_0$ phase in which $C_\pm^{MF}=0$ at  $q/U_0=0.018$. Both transitions are first order.
        }
\label{Figure4}
\end{figure}
These results are in qualitative agreement with recent observations in three-dimensional lattice.\cite{Liu_2016}

The Mott-superfluid transition is also controlled by the Zeeman parameter $q$ when keeping $t/U_0$ fixed.\cite{Liu_2016}
Figure~\ref{Figure4} shows the successive MI-SF transitions for $t/U_0=0.04$ when increasing $q/U_0$. 
For $q\to -\infty$, the superfluid phase SF$_{\downarrow \uparrow}$ is only composed by $\sigma=\{\downarrow, \uparrow\}$ bosons 
[Fig.~\ref{Figure4}(a)] with 
balanced populations $\rho_+=\rho_-=1$ [Fig.~\ref{Figure4}(b)], whereas for 
$q\to \infty$, the superfluid phase SF$_0$ is only composed by $\sigma=0$ bosons. 
Interestingly enough, the system is in the $\rho=2$ MI phase for $q/U_0\in[-0.08, 0.018]$
with non-integer densities $\rho_\sigma$. This effect is very similar to the one 
observed for  molecular and atomic mixtures with species conversions
for which the gap of the Mott phase is tuned by an extended term in the Hamiltonian.\cite{deforges2015}
In Fig.~\ref{Figure4}, both densities and condensate fractions jump at the transitions, indicating first-order transitions.

\begin{figure}[t]
	\includegraphics[width=1 \columnwidth]{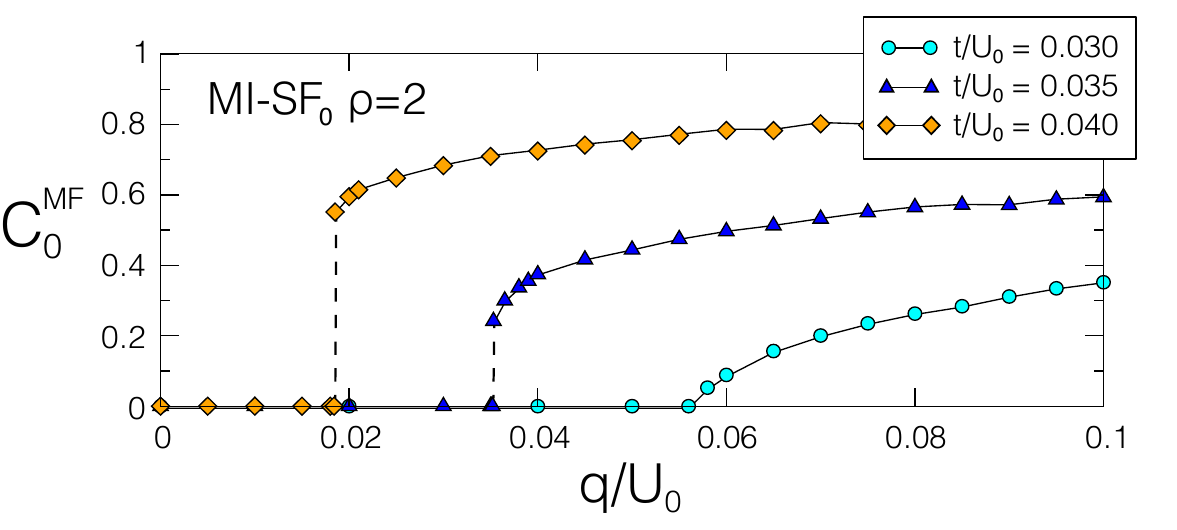}
        \caption {(Color online)  Mean-field MI-SF$_0$ transition controlled by the  $q$. 
        Only the $\sigma=0$ component is condensed in the SF$_0$ phase. The jump in $C^{MF}_0$ is the signal of a first-order transition for 
        $t/U_0=0.040$ and $t/U_0=0.035$, whereas the transition is second order for $t/U_0=0.030$.}
\label{Figure5}
\end{figure}

For $q>0$, the nature of the MI-SF$_0$ transition depends on the ratio $t/U_0$; see Fig.~\ref{Figure5}.
We still observe a first order transition for $t/U_0=0.035$, but the transition becomes second order for $t/U_0=0.030$.
The first order appears at the tip of the Mott lobe where the charge gap is mainly stabilized by the formation of the singlet, i.e., for $t/U_0>0.03$.
Furthermore, the critical $q_c/U_0$ observed in Fig.~\ref{Figure5} increases with $U_0/t$ since the Zeeman term 
has to destroy the energy gap $\Delta \sim U_0$ of the Mott phase.

\section{Quantum Monte Carlo  Phase Diagrams}
\label{sec_4}

We first discuss the phase diagrams, the properties of the phases with respect to the QZE, and then we focus on the quantum phase transitions.

\subsection{Phase Diagrams}
\label{sec4_subA}

For a fixed hopping $t/U_0$, the boundaries of the $\rho$ Mott lobe are calculated with
$\mu^+(\rho)=E(\rho L^2+1)-E(\rho L^2)$ and $\mu^-(\rho)=E(\rho L^2)-E(\rho L^2-1)$.
The charge gap $\Delta \equiv \mu^+-\mu^-$ vanishes at the tip of the Mott lobe in the thermodynamic limit. 
\begin{figure*}[t]
\hbox{\includegraphics[width=1 \columnwidth]{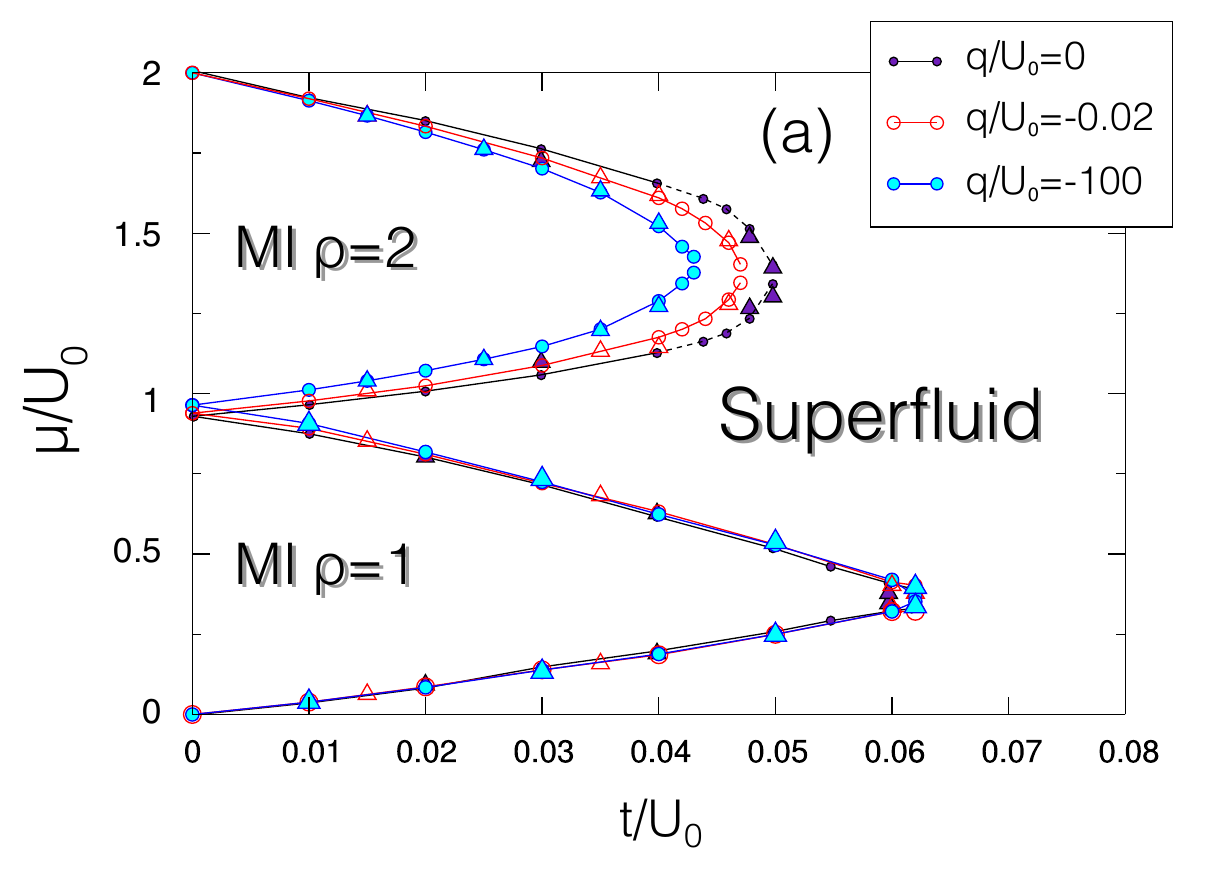}
  \includegraphics[width=1 \columnwidth]{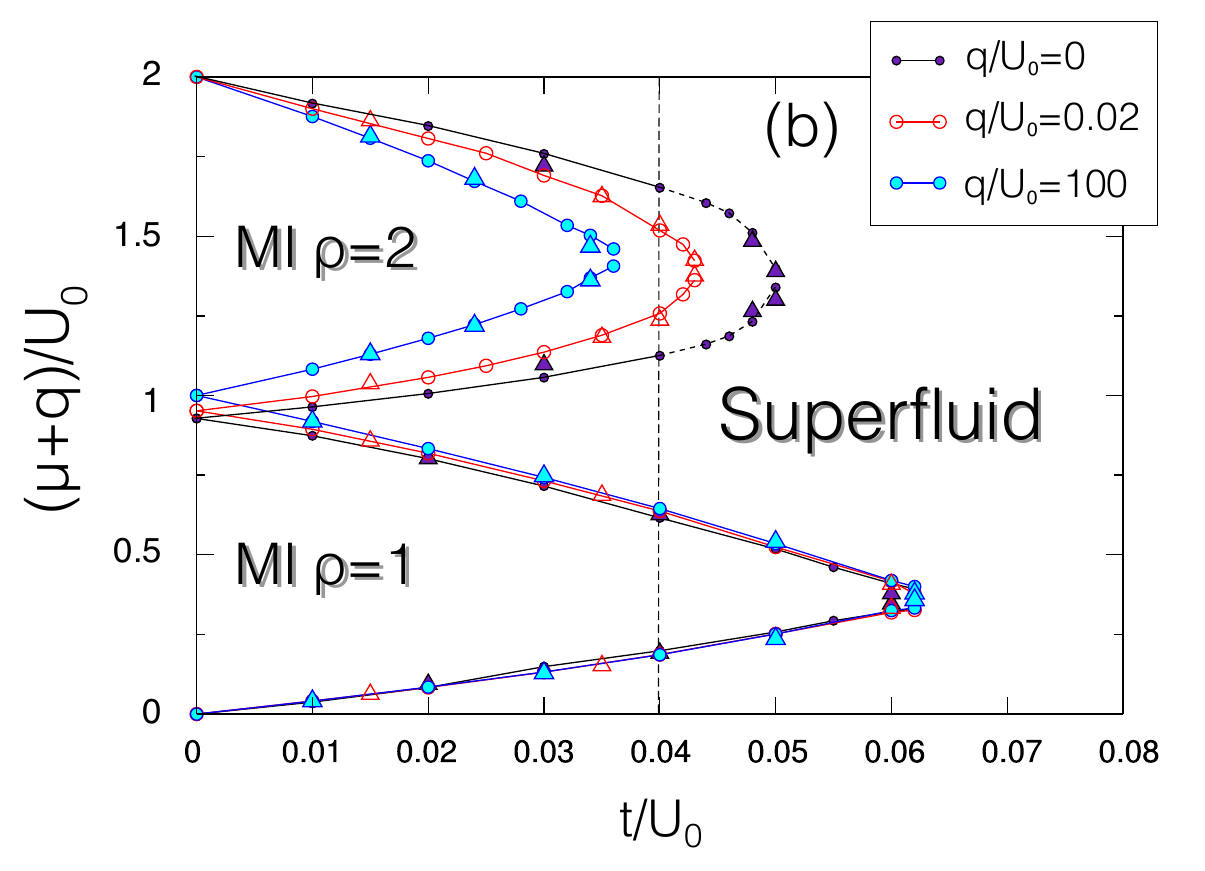}}
        \caption {(Color online) QMC phase diagrams with 
        $U_2/U_0=0.036$ for (a) negative $q$ and (b) positive $q$ values (circles: $L=8$; triangles: $L=12$). 
        Contrary to the $\rho=1$ Mott lobe, $q$ strongly affects the tip of the $\rho=2$ Mott lobe.
        The dashed (plain) lines indicate a first- (second-) order transition.
        Contrary to the mean-field results Fig.~\ref{Figure2}, all the transitions are continuous for $q\neq0$.
        The vertical dashed line in (b) at $t/U_0=0.04$ corresponds to the cut shown in Fig.~\ref{Figure18}. 
        }
        \label{Figure6}     
\end{figure*}
The QMC phase diagrams are plotted in Fig.~\ref{Figure6} for positive and negative $q$. 
Similar to the mean-field phase diagrams, Fig.~\ref{Figure2}, 
the tip of the $\rho=1$ Mott lobe does not vary with $q$, contrary to the one of the $\rho=2$ MI phase.
This is because the charge gap at the tip of the $\rho=2$ Mott lobe -- stabilized by the 
creation of pairs of bosons in the singlet state -- is destroyed by $q$, thus leaving the space to the superfluid phase when increasing $|q|$.
As compared to the mean field, which underestimates the quantum fluctuations, 
the MI lobes end at larger $t_c/U_0$ values, for all $\rho$ and $q$.
Therefore, the MI regions are much bigger than the mean field ones, as expected, 
and the $\rho=1$ Mott phase ends at $t_c/U_0\simeq 0.06$, in agreement with the  
single-species Bose-Hubbard model.\cite{Capogrosso_Sansone_2008}  
In all the phases, we observe neither ferromagnetism, nor N\'eel order, 
i.e., no peak in magnetic structure factor $\mathcal S_{\alpha \alpha}({\bf k})$ [Eq.~(\ref{magnetic_structure_factor})] for $k=0$ and $k_{x,y}=\pi$.
Nevertheless, we do observe  a signal of nematic order, as discussed below.

\subsection{Densities, nematic order and MI-SF transition vs. QZE}
\label{sec4_subB}

\begin{figure}[h]
	\includegraphics[width=1 \columnwidth]{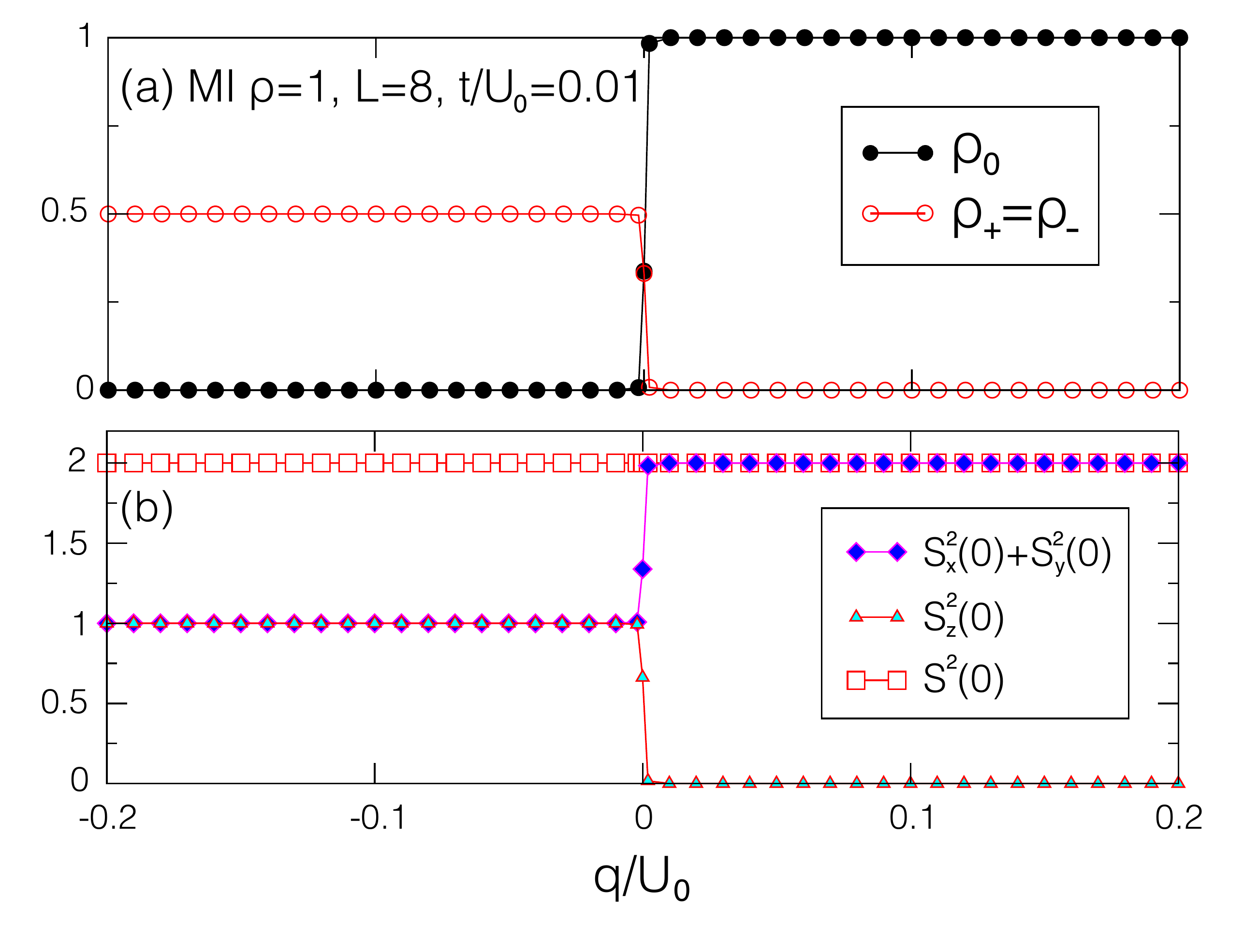}
        \caption {(Color online) QMC data in the $\rho=1$ MI phase. 
        (a) The Zeeman term  $q$  allows one to tune the populations
        from a state fully composed by  $\sigma=\{\downarrow, \uparrow\}$ particles for $q<0$ to a full $\sigma=0$ state  for  $q>0$.
        (b) As a consequence, $q$ affects the components of the magnetic local moment, whereas its amplitude remains fixed $S^2(0)=2$.}
        \label{Figure7}
\end{figure}
We first focus on the magnetic properties of the $\rho=1$ Mott phase  when varying $q$.
Figure~\ref{Figure7}(a) shows the effect
of the Zeeman term on the populations at fixed  $t/U_0=0.01$. As expected, for $q<0$, the system is only composed by particles in 
states  $\sigma=\{\downarrow, \uparrow\}$  ($\rho_\pm=1/2$), whereas the system is fully composed by particles in state $\sigma=0$ ($\rho_0=1$)
for $q>0$. For $q=0$, the populations are balanced $\rho_0=\rho_\pm \simeq1/3$.
Nevertheless, for all $q$, the local magnetic moment remains fixed at $S^2(0)=2$; see Fig.~\ref{Figure7}(b). 
Therefore, the magnetic term in the Hamiltonian is constant and does not compete with the Zeeman term. 
Since $q$  affects the populations, the components of the local magnetic moment are also affected, see Eq.~(\ref{magnetic_local_moment_squared}),  such 
that $S^2_{x, y}(0)=\rho_\pm,  ~S^2_z(0)=2\rho_\pm$ for $q<0$,  and $S^2_{x, y}(0)=\rho_0, S^2_z(0)=0$ for $q>0$, and
the local magnetic moment saturates to its maximal value $S^2(0)=2\rho, \forall q$.\cite{deforges2014} 
A finite local magnetic moment is a necessary but not sufficient condition for the establishment of a magnetic ordering, and 
we should carefully look at the nematic correlation functions.\cite{DeChiara_2011}
\begin{figure}[t]
	\includegraphics[width=1 \columnwidth]{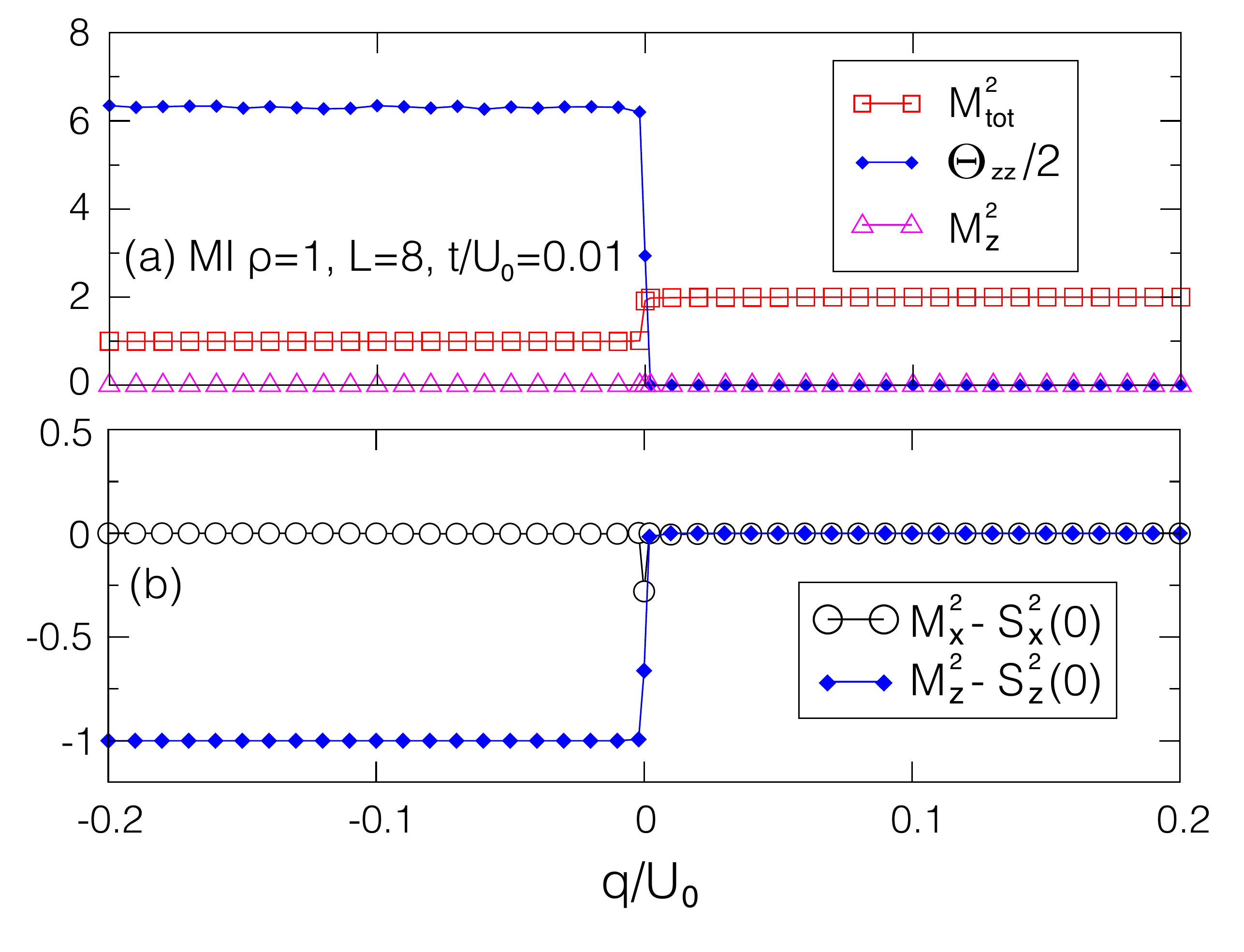}
        \caption {(Color online) QMC data in the $\rho=1$ MI phase. 
        (a) $\Theta_{zz}\neq0$ and $M^2_z=0$ are indicators of a nematic order along $z$.
        (b) The nematic order along $z$ implies  $M^2_{z}-S_z^2(0) <0$. 
        In the $xy$ plane, we observe nonvanishing correlation $M^2_{x,y}-S_{x,y}^2(0) \neq0$ only for $q=0$.}
        \label{Figure8}
\end{figure}
The magnetic correlation functions are plotted in Fig.~\ref{Figure8}.
The nematic order parameter  $\Theta_{zz}\neq0$ is non zero for $q\leq0$; see
 Fig.~\ref{Figure8}(a).
Indeed, a nematic order along $z$ is consistent with a vanishing magnetization $M^2_z=0$,\cite{deforges2014, Katsura}
and  non vanishing correlation  $M^2_{z}-S_z^2(0) =  \frac{1}{L^4} \sum_{ \bf r, R\neq 0} \langle   {\hat S}_{z, {\bf r}} {\hat  S}_{ z, {\bf r+R}} \rangle \neq 0$ 
observed in Fig.~\ref{Figure8}(b). 
In the $xy$ plane, we only observe non vanishing correlation functions for $q=0$ for which $M^2_{x,y}-S_{x,y}^2(0) \neq0$.
In conclusion, we observe a nematic phase for $q\leq0$.

\begin{figure}[t]
	\includegraphics[width=1 \columnwidth]{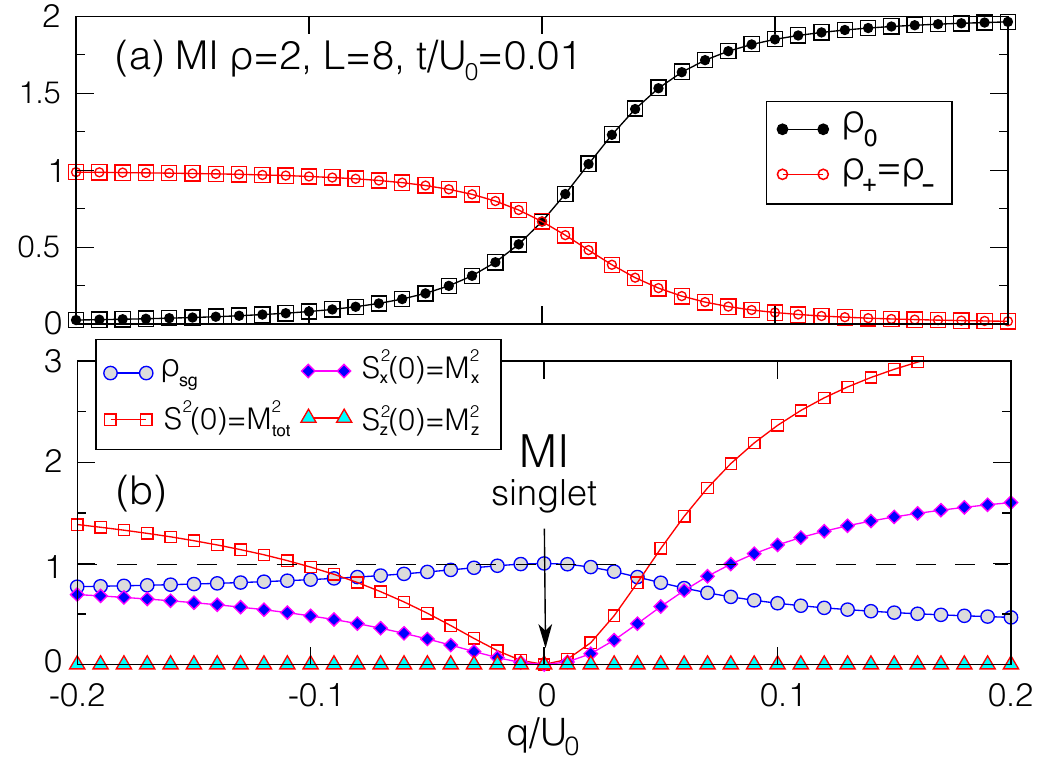}
        \caption {(Color online) QMC data in the $\rho=2$ MI phase. (a)  The Zeeman term  $q$  tunes the populations
        from a state fully composed by $\sigma=\{\downarrow, \uparrow\}$ particles for $q\to -\infty$ to a full $\sigma=0$ state  for  $q\to +\infty$.
        For $q=0$, the system  adopts a singlet state with balanced populations. 
        Squares are the densities calculated using the local wave function Eq.~(\ref{Local_wave_function_rho2}).
        (b) The magnetic correlations are absent since  $M^2_{\rm tot}-S^2(0)=0,  \forall q$.}
\label{Figure9}
\end{figure}

The situation is very different in the $\rho=2$ Mott phase since the local magnetic moment is  fully minimized, i.e., $S^2(0)=0$, in the $t/U_0\to0$ limit.
Therefore,  the local magnetic moment $S^2(0)$ and the Zeeman term are in competition for minimizing the 
free energy of Hamiltonian Eq.~(\ref{Hamiltonian_compact_form}).
Similarto Fig.~\ref{Figure7}(a), the system is only composed by particles in 
states $\sigma=\{\downarrow, \uparrow\}$  ($\rho_\pm=1$) for $q/U_0 \to -\infty$,  whereas the system is fully composed by particles in state $\sigma=0$ ($\rho_0=2$)
for $q/U_0 \to +\infty$; see Fig.~\ref{Figure9}(a).
Nevertheless, contrary to the $\rho=1$ case plotted in Fig.~\ref{Figure7}(a), 
we observe a smooth crossover at $q=0$ for which  the system adopts a singlet state such that $\rho_{sg}=1$ and $S^2(0)=0$;
see Fig.~\ref{Figure9}(b). 
For $q\neq 0$, the minimization of the Zeeman term breaks the singlet state which leads to a 
non vanishing local moment $S^2(0)\neq0$ in the plane $xy$, i.e., $S_{x,y}^2(0)\neq0$, but  with $S_z^2(0)=0$.
These local quantities are well described by the  on-site wave function Eq.~(\ref{Local_wave_function_rho2}).
Clearly, $  |\Phi_{\rho=2, q\to -\infty}\rangle =  |1,0,1\rangle$ leads to $\rho_\pm=1$, and $ |\Phi_{\rho=2, q\to +\infty}\rangle =  -|0,2,0\rangle$ leads to $\rho_0=2$.
For $q=0$, the singlet state $|\Phi_{\rho=2, q=0}\rangle= \frac{1}{\sqrt{3}} \left(   \sqrt{2} |1,0,1\rangle  -|0,2,0\rangle  \right)$ leads to $\rho_\pm=\rho_0=2/3$
and $ \langle a_{0 {\bf r}}^\dagger      a_{0 {\bf r}}^\dagger       a_{\uparrow {\bf r}}^{\phantom\dagger}        a_{\downarrow {\bf r}}^{\phantom\dagger} \rangle =-2/3$
from Eq.~(\ref{magnetic_local_moment_squared}).
The finite local moment obtained for $q\neq0$ may suggest a magnetic ordering in the 
$xy$ plane, since  $S^2_{x, y}(0)-\frac{1}{3}S^2(0) \neq 0$.\cite{Imambekov_2004, demler02} 
In fact, the contribution of the spin-spin correlation functions, when removing the auto-correlation contribution,
vanishes, i.e., $M^2_{\rm tot}-S^2(0)= \frac{1}{L^4} \sum_{\sigma, \bf r, R\neq 0} \langle   {\hat S}_{\sigma, {\bf r}} {\hat  S}_{ \sigma, {\bf r+R}} \rangle=0$.
Therefore, there is no magnetic order for $t/U_0=0.01$ in the $\rho=2$ Mott phase $\forall q$.

\begin{figure}[t]
	\includegraphics[width=1 \columnwidth]{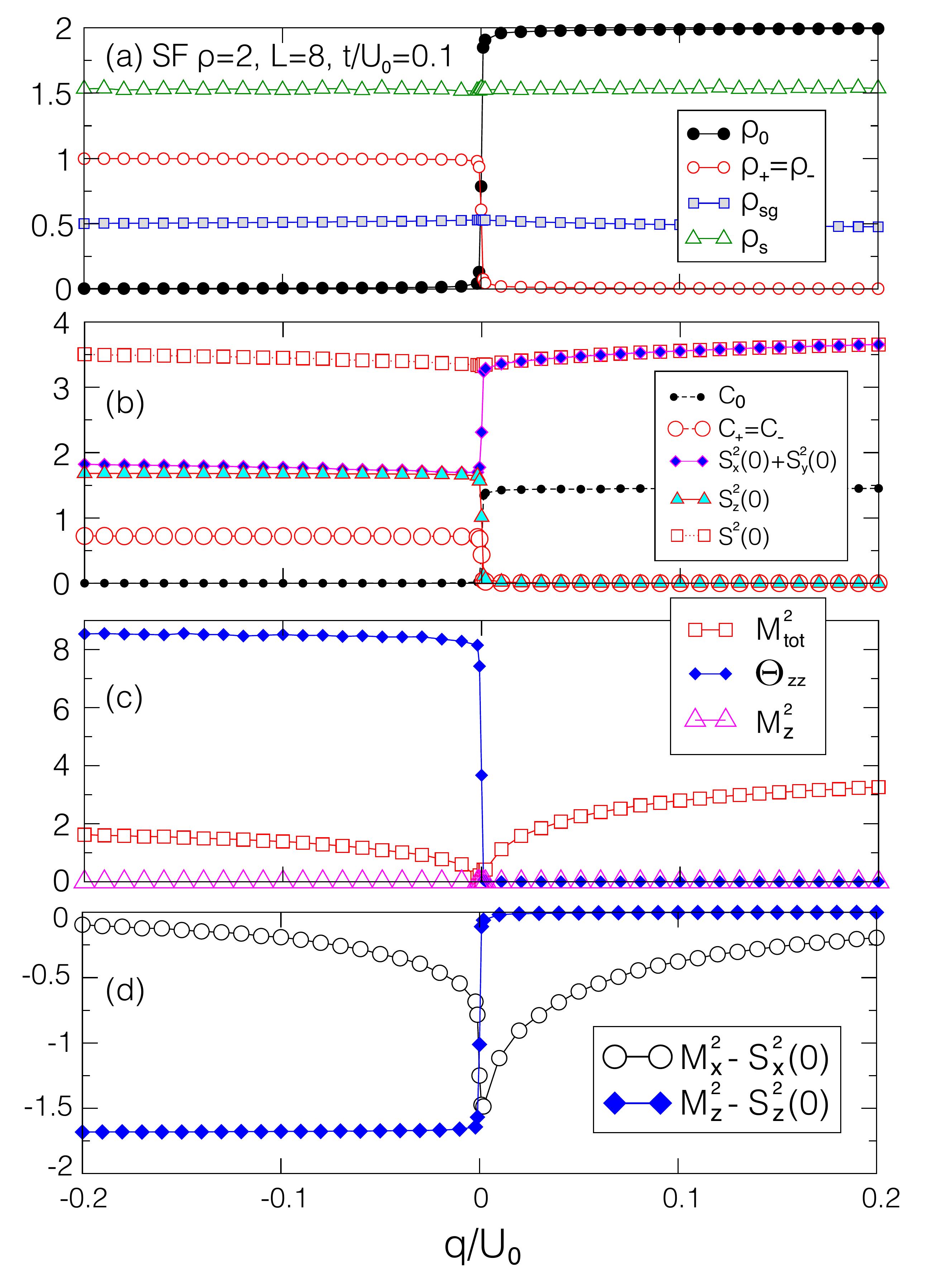}
        \caption {(Color online) 
        QMC data in the $\rho=2$ SF phase with $L=8$: (a) the superfluid and singlet densities $\rho_s$, $\rho_{sg}$ 
        are insensitive to $q$. (b) The condensate is  supported by $\sigma=\{\downarrow, \uparrow\}$ ($\sigma=0$) bosons for $q<0$ ($q>0$) 
        and the $z$ component of local magnetic  moment $S_z^2(0)$ vanishes for  $q>0$ due to $\rho_\pm=0$.
        (c),(d) The magnetic moment $M^2_{\rm tot}$ vanishes for $q=0$ and 
        we observe a  nematic order along $z$ only for  $q\leq0$  ($\Theta_{zz}\neq0$) associated with negative correlations $M^2_{z}-S_z^2(0)<0$.
        In the $xy$ plane, the amplitude of the spin-spin correlations $|M^2_{x, y}-S_{x, y}^2(0)|$ reaches its maximum for  $q=0$
        and vanishes for $q\to \pm \infty$.
        }
\label{Figure10}
\end{figure}
We now focus on the $\rho=2$ superfluid phase, in which the kinetic term is dominant:
the Bose-Einstein condensation -- associated with the spontaneous U(1) symmetry
breaking -- occurs; hence $\rho_s\neq0$ and $C_\sigma\neq0$. 
Therefore, the QZE also affects the condensate populations $C_\sigma$.
As an example, we focus on the case with $t/U_0=0.1$; see Fig.~\ref{Figure10}.
The superfluid density $\rho_s$ and singlet density $\rho_{sg}$ do not significantly vary with $q$, whereas the populations
evolve from a state fully composed by $\sigma=\{\downarrow, \uparrow\}$ particles for $q\to -\infty$ to a full $\sigma=0$ state  for  $q\to +\infty$ [Fig.~\ref{Figure10}(a)].
Note that the redistribution of the population at $q=0$ is sharp -- similar to the $\rho=1$ MI phase  in
Fig.~\ref{Figure7}(a).
This is because the spin-spin interaction term in Eq.~(\ref{Hamiltonian_compact_form}) is now in competition with a dominant kinetic term
and has a smaller effective strength.
The condensates $C_\sigma$ naturally follow the densities $\rho_\sigma$ [Fig.~\ref{Figure10}(b)].
The components of the local magnetic moment behave qualitatively in the same way as in the $\rho=1$ Mott phase, 
Fig.~\ref{Figure10}(b), but 
do not reach their maximal values since a small fraction of particles remains in the singlet state  for $t/U_0=0.1$, that is  
$S^2_{x, y}(0)<\rho_\pm,  ~S^2_z(0)<2\rho_\pm$ for $q<0$,  and $S^2_{x, y}(0)<\rho_0$ for $q>0$; hence $S^2(0)<2\rho, \forall q$\cite{deforges2014} 
(or equivalently $ \langle a_{0 {\bf r}}^\dagger      a_{0 {\bf r}}^\dagger       a_{\uparrow {\bf r}}^{\phantom\dagger}        a_{\downarrow {\bf r}}^{\phantom\dagger} \rangle <0$).
Compared to the $\rho=1$ Mott phase, 
the director of the nematic order in the $\rho=2$ superfluid can belong to the three axes in a larger range of  $q$:
the signal of a nematic phase along $z$ is clearly observed in Fig.~\ref{Figure10}(c)
where $\Theta_{zz}\neq0$ for $q\leq 0$.
This statement is strengthened  by the non vanishing correlation function
$M^2_{z}-S_z^2(0) <0$ in Fig.~\ref{Figure10}(d). 
Nevertheless, $\Theta_{zz}=0$ for $q> 0$ indicates a vanishing nematic order along $z$.
Furthermore, in the $xy$ plane, the amplitude of the spin-spin correlations $|M^2_{x, y}-S_{x, y}^2(0)|$ reaches its maximum for  $q\simeq0$
and vanishes for $q\to \pm \infty$. Interestingly enough, the nematic correlations $|M^2_{\rm tot}-S^2(0)|$ are maximized when the 
global magnetism $M^2_{\rm tot}$ vanishes at $q=0$ [Fig.~\ref{Figure10}(c)]. 
Our results are in good agreement with previous studies\cite{Katsura, Deforges_2013, deforges2014}
which have predicted a nematic order in the superfluid phase with broken SU(2) symmetry.
In the limit $U_2\ll t $, with $U_0=q=0$ and $\rho=2$, the nematic correlations take the exact value $M^2_{\rm tot}-S^2(0)=-2\rho=-4$.\cite{deforges2014}
With the data of Fig.~\ref{Figure10}, we obviously 
find a smaller value $| M^2_{\rm tot}-S^2(0)| \simeq 3.3$ because of the non-negligible interactions $U_0=10~t$ which allow the formation 
of a small fraction of particles in the singlet state,
hence $\langle a_{0 {\bf r}}^\dagger      a_{0 {\bf r}}^\dagger       a_{\uparrow {\bf r}}^{\phantom\dagger}        a_{\downarrow {\bf r}}^{\phantom\dagger} \rangle < 0$ 
from Eq.~(\ref{magnetic_local_moment_squared}), thus partially destroying the nematic order.
In conclusion, the director of the nematic order evolves form a director along $z$ for $q\to -\infty$ to a director with finite $xyz$ components at $q\simeq0$.
For $q>0$, the director belongs to the $xy$ plane and the nematic order disappears as $q$ is increased.

\begin{figure}[t]
	\includegraphics[width=1 \columnwidth]{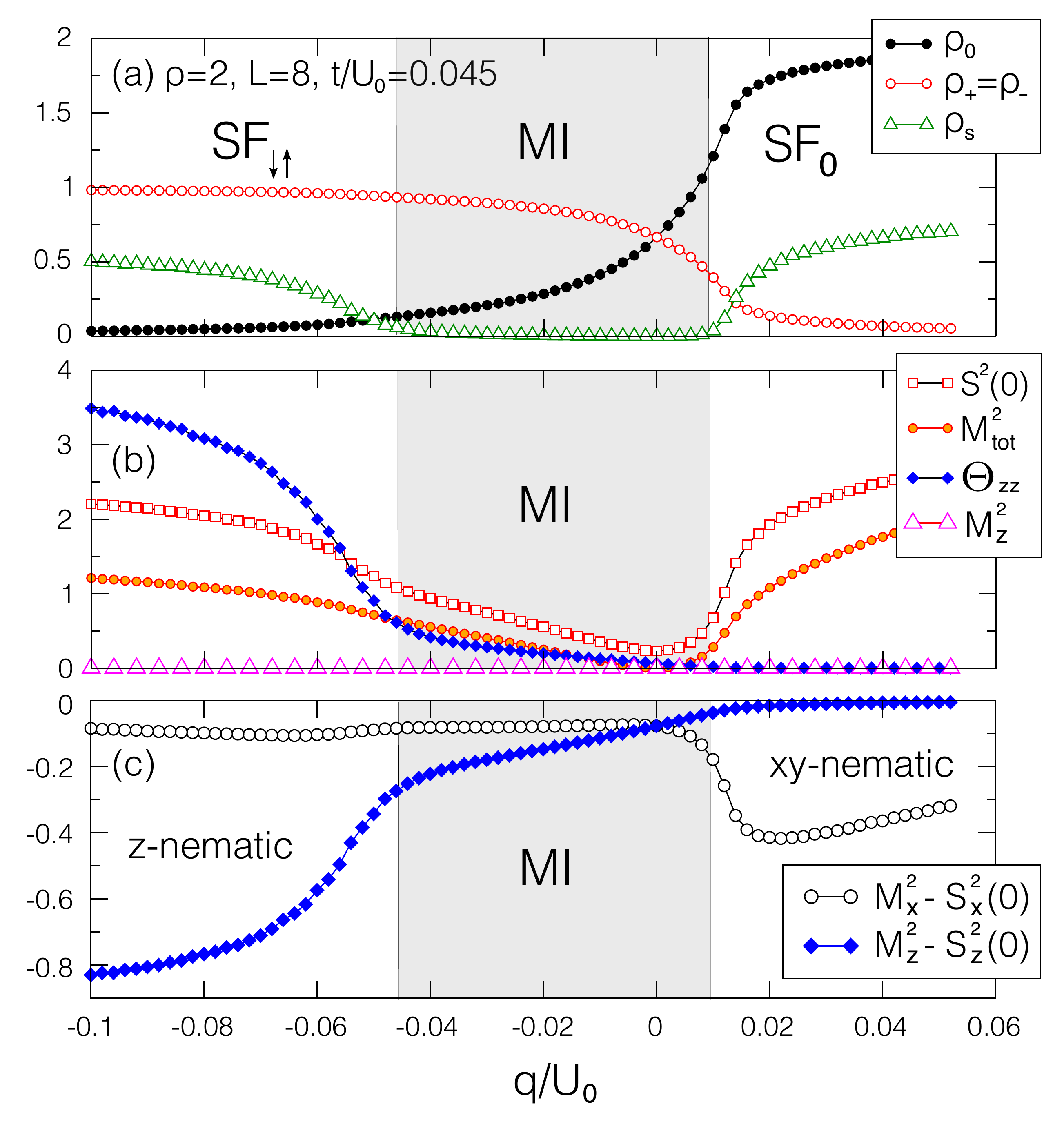}
        \caption {(Color online) QMC data for $\rho=2$ and $L=8$ at fixed hopping $t/U_0=0.045$: $q$ acts as a control parameter for the MI-SF transition.
        For $q\to -\infty$ the system is in the SF$_{\downarrow \uparrow}$ with a nematic director along $z$ axis and enters in the MI phase at $q/U_0 \simeq-0.045$ when increasing
        $q$. Then, the system continuously adopts a SF$_0$ phase with a nematic director belonging to the $xy$ plane at $q/U_0 \simeq 0.01$. Both transitions are second order.
        }
\label{Figure11}
\end{figure}

For intermediate $t/U_0$, the QZE can act as a control parameter of the Mott-superfluid transition as shown at the mean field level in Fig.~\ref{Figure4}.
This effect is confirmed by our QMC simulations for $t/U_0=0.045$, 
where the system adopts a Mott phase for $q/U_0\in[-0.045,0.01]$ with $\rho_s=0$ and is superfluid $\rho_s\neq0$ otherwise;
see Fig.~\ref{Figure11}(a).
In agreement with the mean-field results, the densities evolve from $\rho_\pm=1$ for $q/U_0\to -\infty$ to $\rho_0=2$ for $q/U_0\to +\infty$, 
and $\rho_\pm=\rho_0=2/3$ at $q/U_0=0$ where the  local magnetic moment $S^2(0)$ and the magnetization $M^2_{\rm tot}$ are minimized; see Fig.~\ref{Figure11}(b).
Although the magnetization is strictly zero along the $z$ axis, i.e., $M^2_z=0$, we observe a clear signal of a nematic order along the $z$ axis in the 
SF$_{\downarrow \uparrow}$ phase, i.e., $\Theta_{zz}\neq0$; see Fig.~\ref{Figure11}(b).
Furthermore, we observe nematic correlation $M^2_{z}-S_z^2(0) <0$ in the SF$_{\downarrow \uparrow}$ phase, Fig.~\ref{Figure11}(c).
In the SF$_0$ phase for $q/U_0 > 0.01$, the nematic order is developed in the  $xy$ plane, where 
$M^2_{x, y}-S_{x, y}^2(0)<0$; see Fig.~\ref{Figure11}(c).
Therefore, the QZE allows one to control both the phase coherence and  the director of the
nematic order.
Contrary to the mean-field predictions, our data for $L=8, 10$ show continuous transitions.

\subsection{Nature of the $\rho=1, 2$  MI-SF transitions}
\label{sec4_subC}

Our mean-field results, Fig.~\ref{Figure3},  suggest that the nature of the MI-SF transition varies with $q$ for $\rho=2$, whereas it remains 
second order for $\rho=1$.
We now use the QMC method for investigating this effect.

\begin{figure}[t]
	\includegraphics[width=1 \columnwidth]{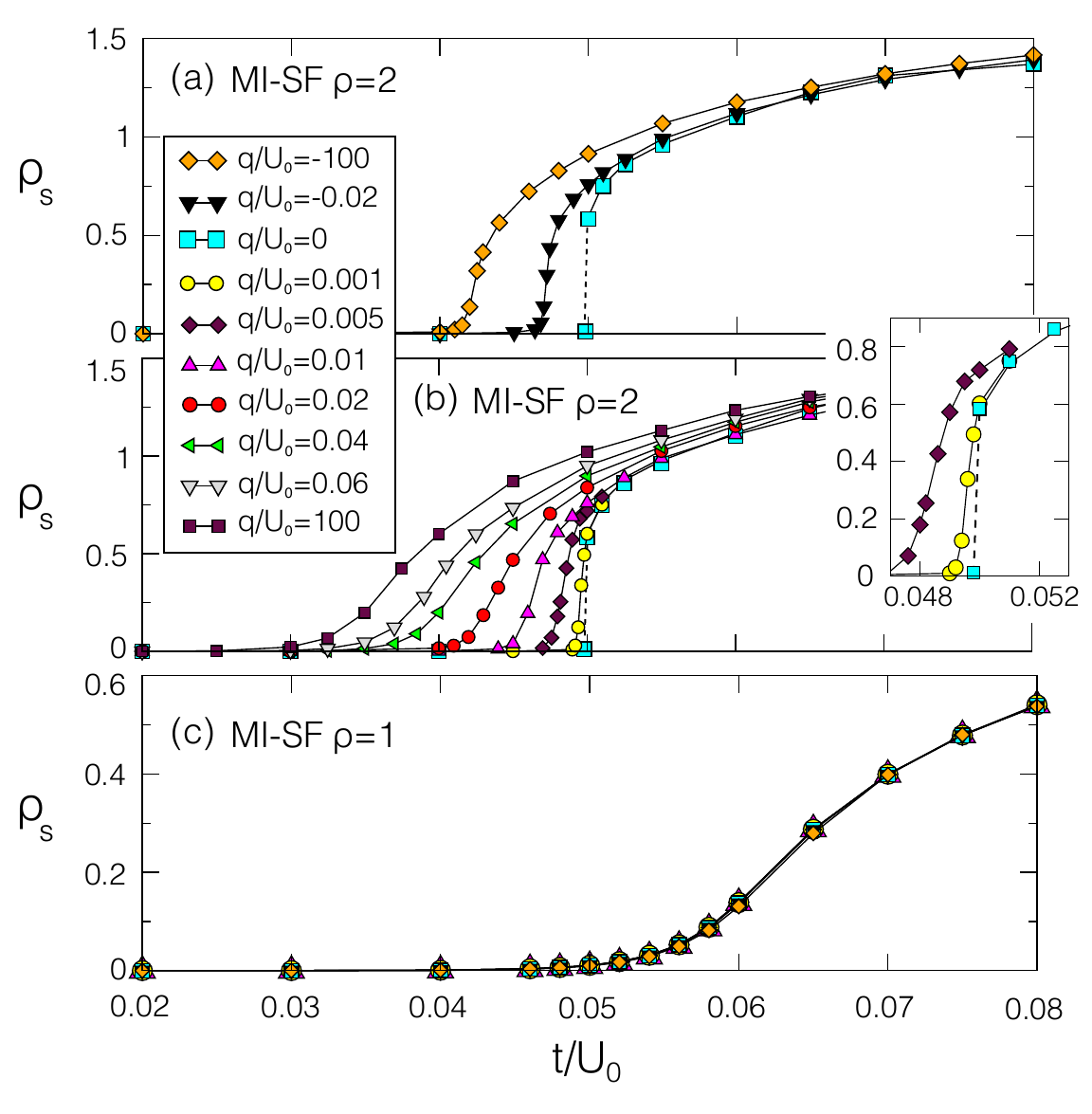}
        \caption {(Color online)  
         QMC data of the superfluid density $\rho_s$ at the MI-SF transition with $L=8$  for  $\rho=2$  with (a) negative $q$  and (b) positive $q$, and (c) for $\rho=1$, $\forall q$.
         Contrary to the $\rho=2$ case, $\rho_s$  does not depend on $q$ for $\rho=1$.
         We only observe a discontinuity in $\rho_s$ for $\rho=2$ and  $q=0$, and the transition is first order; see zoom inset panel (b).}
\label{Figure12}
\end{figure}

Similar to Fig.~\ref{Figure3}, the critical hopping $t_c/U_0$ at the transition is highly sensitive to $q$
for $\rho=2$, but does not change for $\rho=1$; see  Fig.~\ref{Figure12}.
These different behavior comes from the possible minimization of $S^2(0)$ for $\rho=2$, whereas $S^2(0)=2, \forall q$ for $\rho=1$.
Therefore,  the minimization of the free energy of the Hamiltonian Eq.~(\ref{Hamiltonian_compact_form}) 
leads to a competition between the Zeeman term and the minimization of $S^2(0)$ only for $\rho=2$.
However, the  QMC and mean-field  predictions are in contradictions:
our QMC simulations explicitly indicate a first-order transition -- signaled by a jump in $\rho_s$ in  Figs.~\ref{Figure12}(a) and \ref{Figure12}(b) --
only for $q/U_0=0$. Indeed, even for very small $q$ values, e.g., $q/U_0=0.001$,   the jump in $\rho_s$ vanishes,  suggesting  a second order transition.
\begin{figure}[t]
	\includegraphics[width=1 \columnwidth]{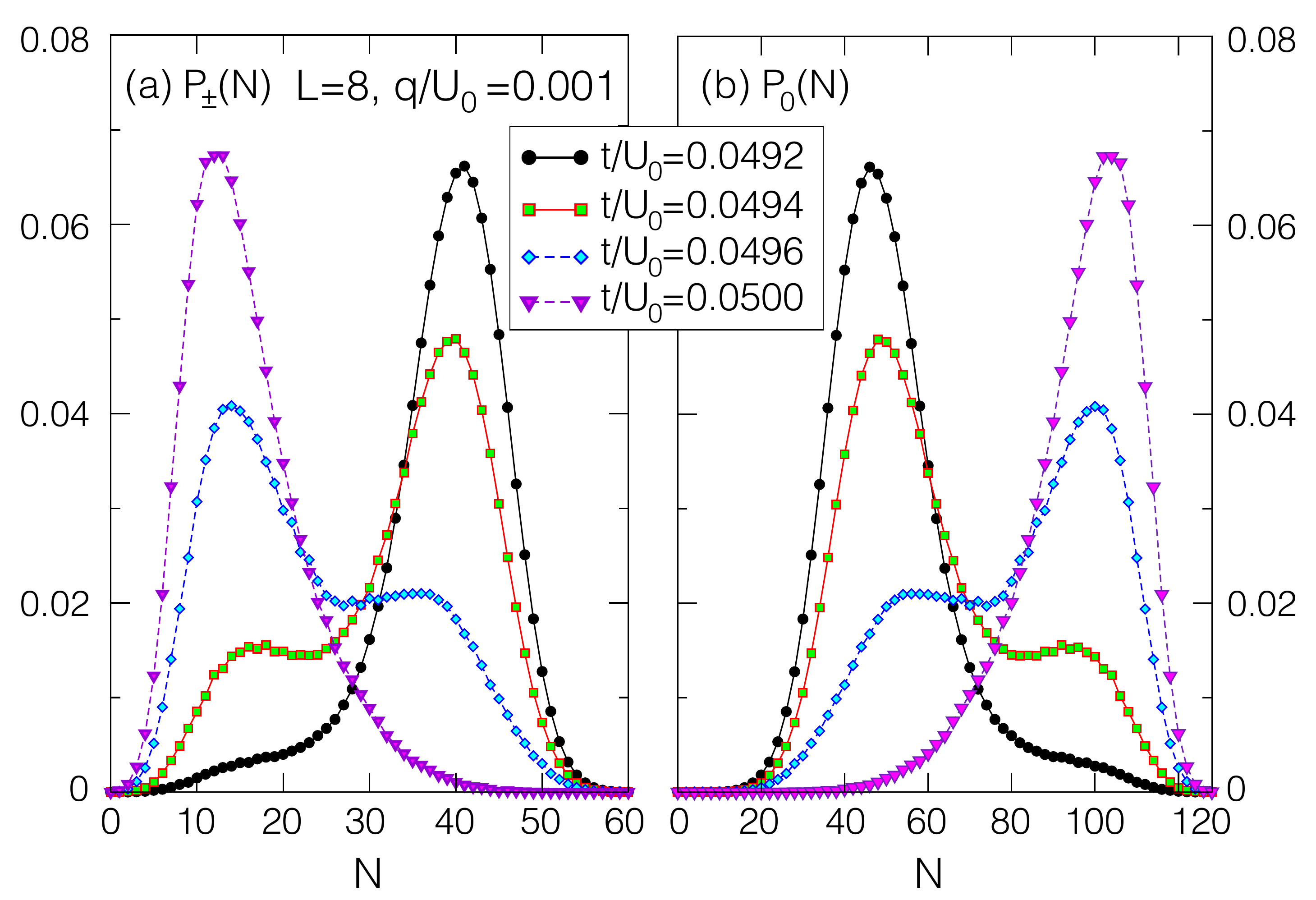}
        \caption {(Color online)  
         QMC data with $L=8$, $\rho=2$, and $q/U_0=0.001$. The density histograms (a) $P_{\pm}(N)$ and (b) $P_{0}(N)$ show a double peak structure close to the transition at 
         $t/U_0=0.0494$ and $t/U_0=0.0496$, thus indicating a weak first-order transition.}
\label{Figure13}
\end{figure}
Nevertheless, the density histograms plotted in Fig.~\ref{Figure13} show double peaks at the transition ($t_c/U_0\simeq0.0494$), 
thus indicating a weak first-order transition for $q/U_0=0.001$.
A similar signal is observed for $q/U_0=-0.005$ but disappears for $q/U_0< -0.005$ and  $q/U_0>0.001$.
Therefore, the transition is found to be first-order only for  $q/U_0 \in[-0.005, 0.001]$.

\begin{figure*}[t]
\hbox{\includegraphics[width=0.5 \columnwidth]{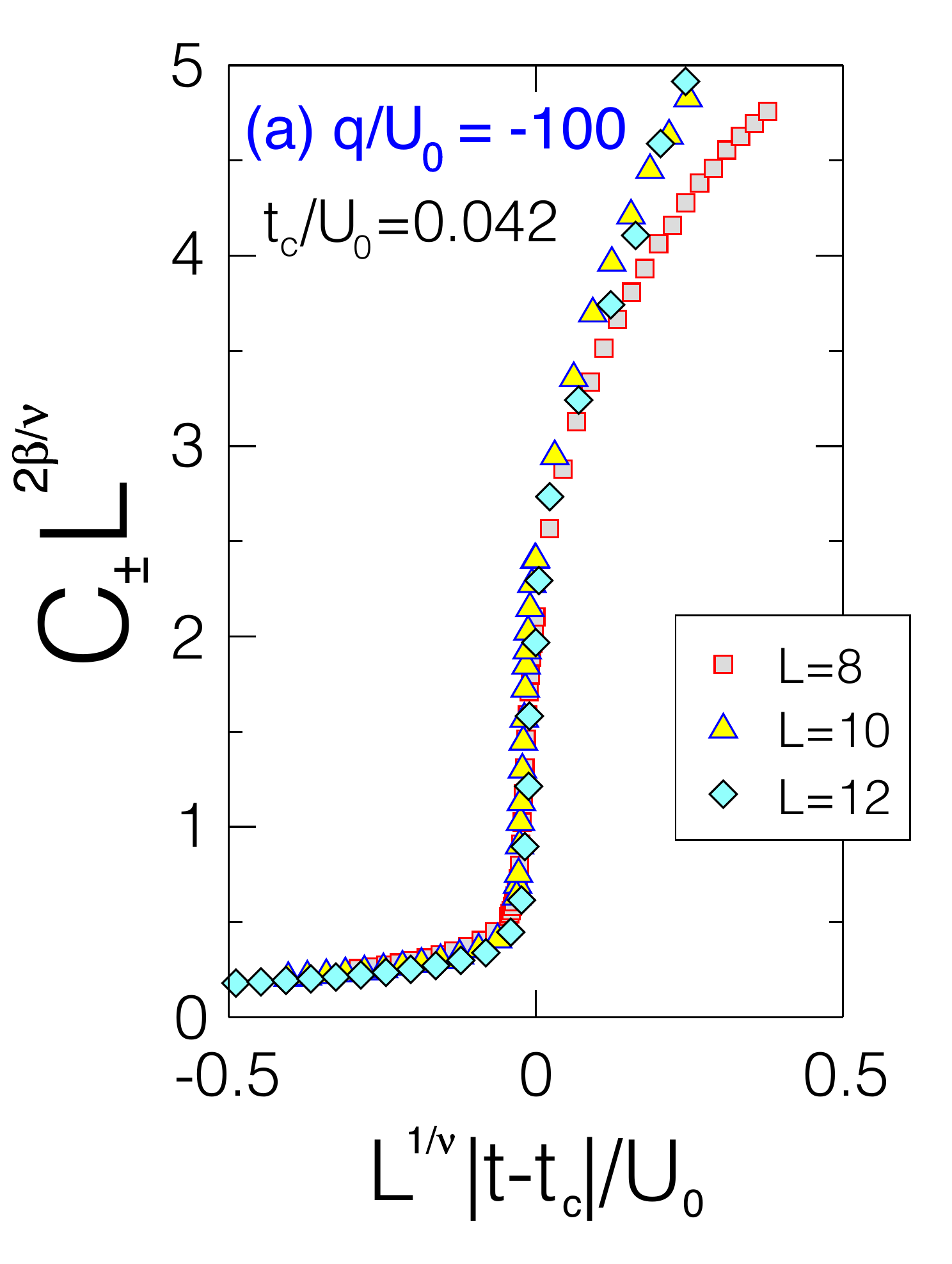}
  \includegraphics[width=0.5 \columnwidth]{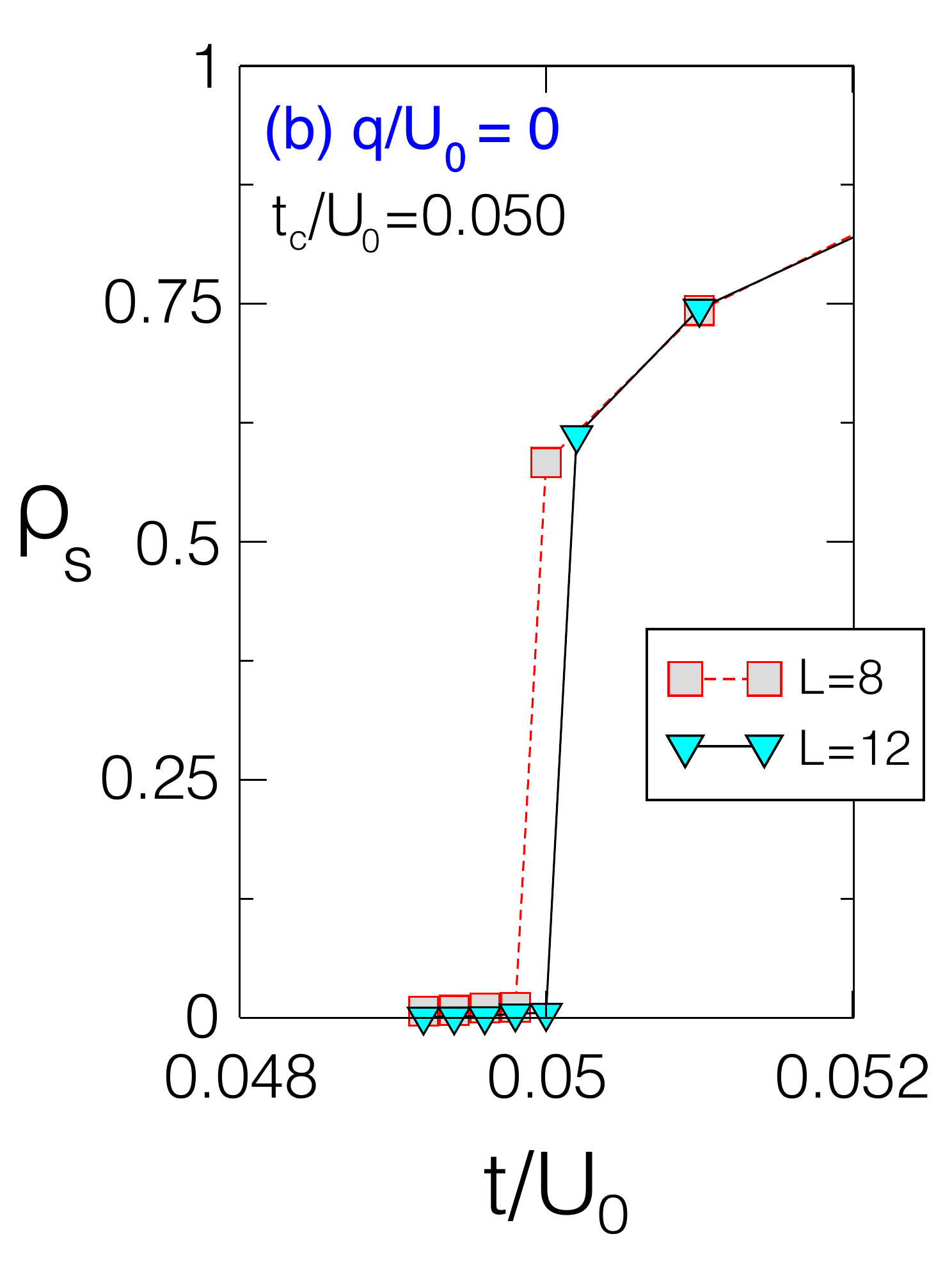}
  \includegraphics[width=0.5 \columnwidth]{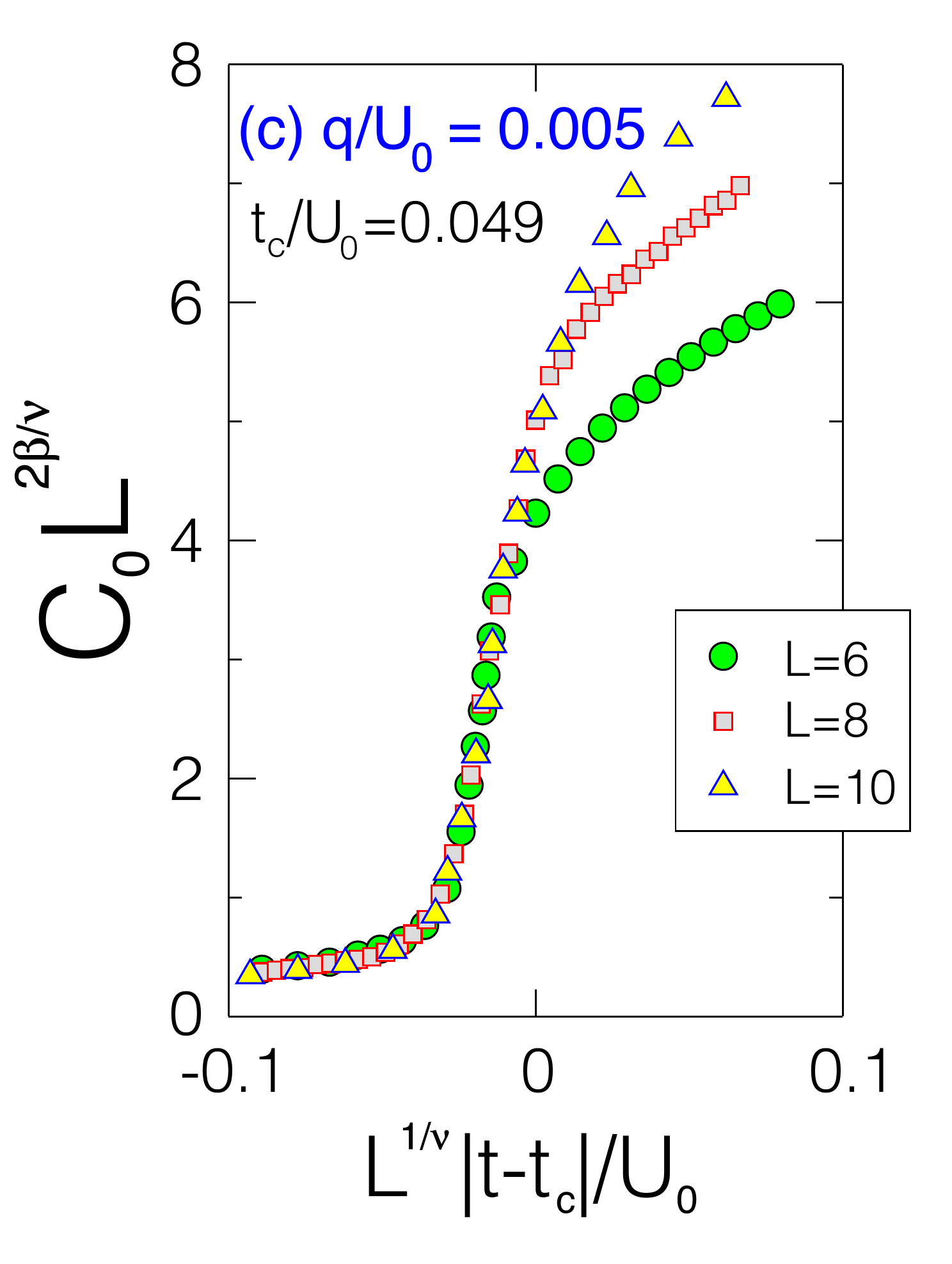}
  \includegraphics[width=0.5 \columnwidth]{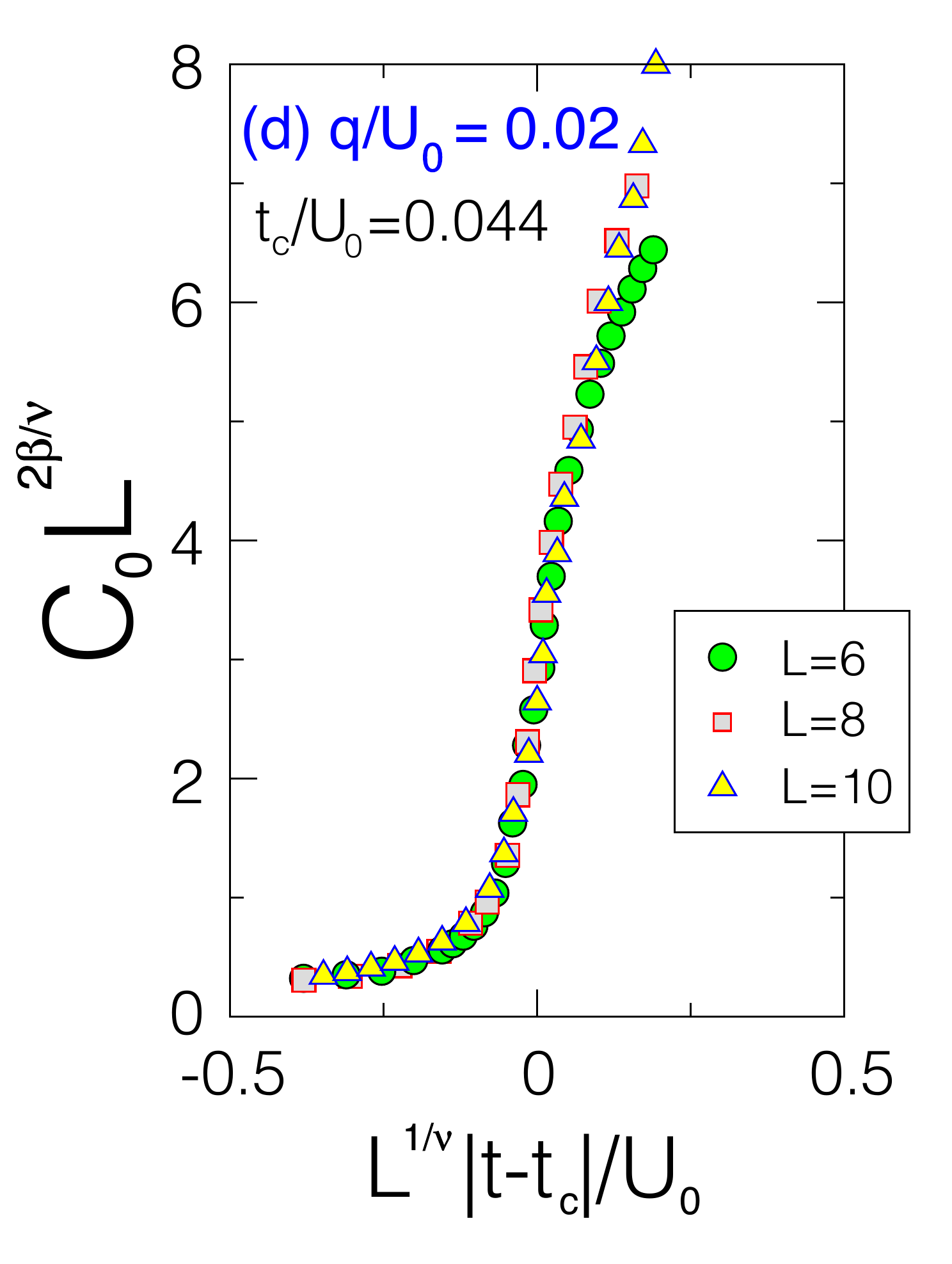}}
        \caption {(Color online)  QMC scaling plots of the $\rho=2$ MI-SF  transition for (a) $q/U_0=-100$, (b) $q/U_0=0$, (c) $q/U_0=0.005$, and (d) $q/U_0=0.02$. 
        For $q\neq0$, the scaling of the condensate fractions $C_0, C_\pm$ are consistent with a transition of the  3D XY nature, even for small  $q/U_0$ (c).        
        Critical exponents of the 3D XY universality classes are used for the scaling, i.e.,  $\beta=0.3479$ and $\nu=0.6706$.\cite{Pelissetto_2002} 
        The jump observed in  $\rho_s$ for $q=0$ (b) indicates a first-order transition.\cite{Deforges_2013, pai08}
         }
        \label{Figure14}     
\end{figure*}

The nature of the $\rho=2$ MI-SF transition is also determined by using finite-size scaling analysis; see Fig.~\ref{Figure14}.
 For $q=0$, the jump in $\rho_s$ remains finite for many sizes; see Fig.~\ref{Figure14}(b),
strengthening the conclusion of a first-order transition associated with the symmetry breaking of both U(1) and SU(2).
This transition has been previously investigated\cite{Deforges_2013, pai08} and other signatures of a first-order transition have been found using QMC simulations 
(e.g., see Fig.~16 of Ref.~\cite{Deforges_2013}).
For  $q/U_0 \notin[-0.005, 0.001]$,  the transition is found to be of the 3D XY nature; see Figs.~\ref{Figure14}(a), \ref{Figure14}(c), and \ref{Figure14}(d).
This result is not surprising in the limit $q/U_0\to +\infty$ since only the component $\sigma=0$ is populated, thus leading to a single species Bose-Hubbard model.
However, this result is more surprising for small positive $q/U_0$ and for large negative $q/U_0$, since a nematic order is established at the MI-SF transition,  thus potentially 
changing the nature of the transition, as observed for $q/U_0=0$.

\subsection{Nematic order at the $\rho=2$ MI-SF transition with fixed $q/U_0$}
\label{sec4_subD}

\begin{figure}[h]
	\includegraphics[width=1 \columnwidth]{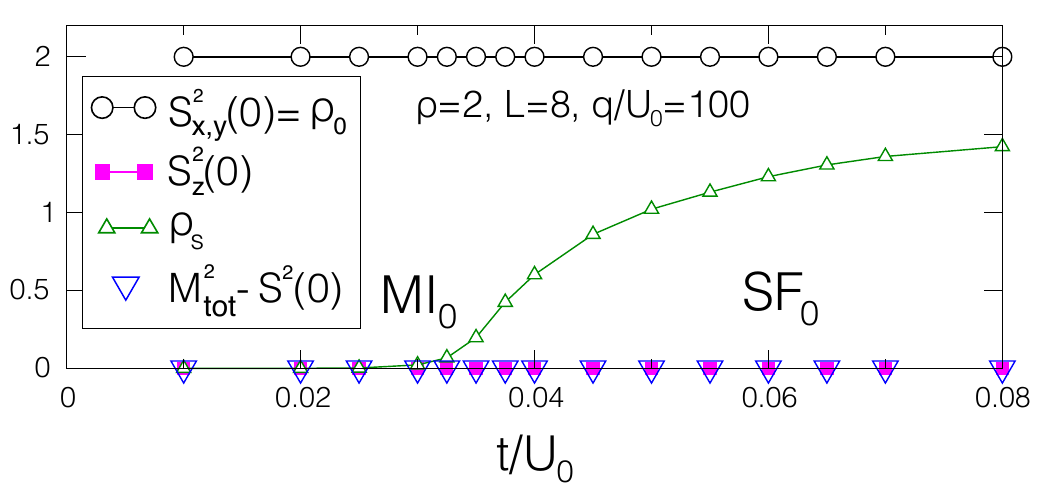}
        \caption {(Color online) QMC data of the $\rho=2$ MI-SF transition with $L=8$  and $q/U_0=100$. 
        The system behaves like the standard single species Bose-Hubbard model with $\rho_0=2$, for which 
        the transition  belongs to the 3D XY universality class,\cite{Fisher_1989}
        without magnetic order $M^2_{\rm tot} -S^2(0)=0$. 
        }
\label{Figure15}
\end{figure}

We now investigate the establishment of the nematic order at the $\rho=2$ MI-SF transition for large and zero $|q|$ values.
To fix the idea, we begin with the simplest case,  $q\to +\infty$,   for which we recover the single species Bose-Hubbard model with $\rho_0=2$ and $\rho_\pm=0$;
see Fig.~\ref{Figure15} for which $q/U_0=100$.
Since $\rho_\pm=0$, the local magnetic moment along $z$ trivially vanishes, $S_{z}^2(0)=0$, whereas $S_{x, y}^2(0)=\rho_0$ is saturated; see
Eq.~(\ref{magnetic_local_moment_squared}).
Obviously, there is no magnetic order,  $M^2_{\rm tot}-S^2(0)=0,  \forall q$.\cite{Blume_Capel}

\begin{figure}[t]
	\includegraphics[width=1 \columnwidth]{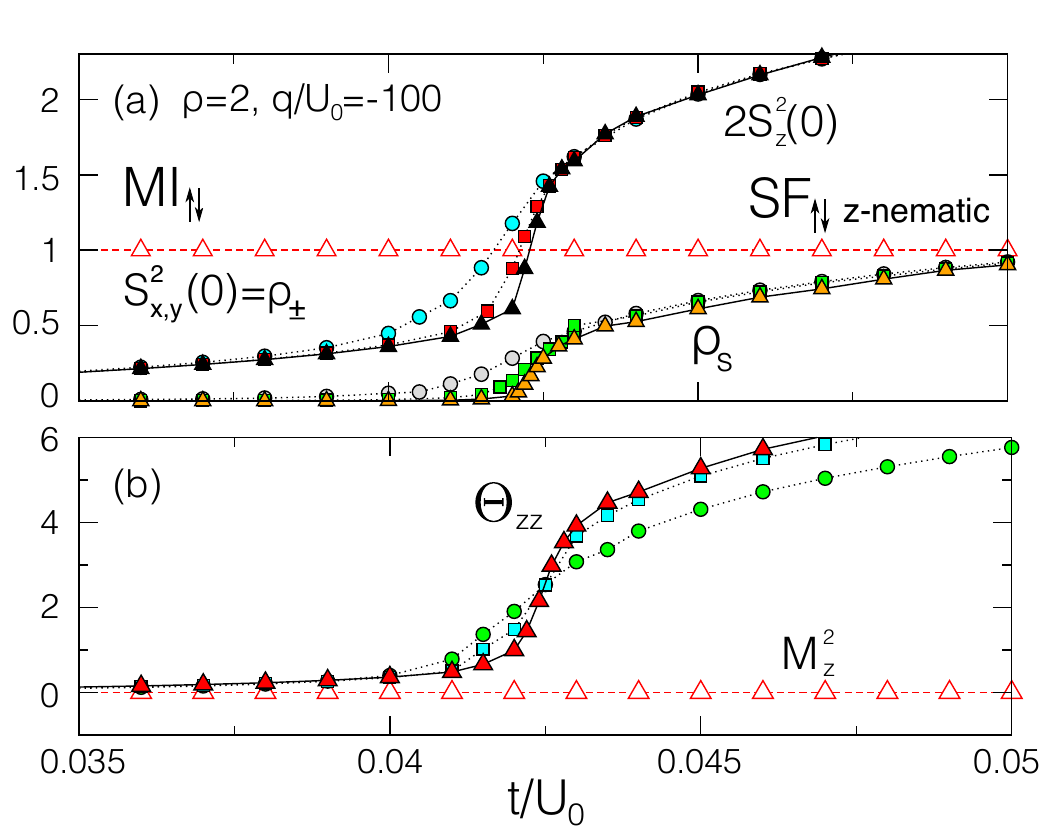}
        \caption {(Color online) QMC data for the $\rho=2$ MI-SF  transition for  $q/U_0=-100$ with $L=6$ (circles), 8 (squares), and 10 (triangles). 
        We observe a continuous transition at $t_c/U_0=0.042$ from a MI$_{\uparrow \downarrow}$ ($\rho_s=0, \Theta_{zz}=0$) 
        to a nematic SF$_{\uparrow \downarrow}$  with a director along $z$ ($\rho_s\neq0, \Theta_{zz}\neq0$).
        }\label{Figure16}
\end{figure}

In the other limit, $q/U_0\to -\infty$, the situation is very different since the establishment of the phase coherence 
leads to the establishment of the nematic order.
In this limit,  the densities read $\rho_\pm=1$ and $\rho_0=0$, $\forall q$, and  the system undergoes a phase transition from
a MI$_{\uparrow \downarrow}$ to a  SF$_{\uparrow \downarrow}$. 
According to Eq.~(\ref{magnetic_local_moment_squared}), this leads to a saturated local magnetic moment in the  $xy$ plane
such that $S^2_{x,y}(0)=\rho_\pm$, whereas a spin degree of freedom remains  along the $z$ axis.
In this case, the magnetic SU(2) symmetry reduces to the  $\mathbb{Z}_2$ Ising symmetry.
For  $t/U_0\to 0$, the Mott phase is described by the on-site wave function $ |\Phi_{\rho=2, q\to -\infty}\rangle =  |1,0,1\rangle$
ensuring  $S_z^2(0)=0$. 
These results are observed for $q/U_0=-100$ in Fig.~\ref{Figure16} (a).
When increasing $t/U_0$, the $z$ component of the magnetic local moment $S^2_z(0)$ significantly increases when the phase coherence is established  $\rho_s\neq0$,
at $t_c/U_0=0.042$. Indeed, the phase coherence involves density fluctuations which prevents the full minimization of $S^2_z(0)$.
Nevertheless, the magnetization $M^2_z$  remains zero for all $t/U_0$; see Fig.~\ref{Figure16}(b).
Furthermore, the nematic order parameters $\Theta_{zz}$ becomes finite at $t_c/U_0\simeq0.042$, thus indicating the establishment of a 
nematic order along $z$ associated with  
nematic correlations $M^2_{z}-S_z^2(0) =  \frac{1}{L^4} \sum_{ \bf r, R\neq 0} \langle   {\hat S}_{z, {\bf r}} {\hat  S}_{ z, {\bf r+R}} \rangle \neq 0$ in the superfluid phase.
We observe a scaling of $\Theta_{zz}$ consistent with the 3D Ising universality class,  using  exponents $\beta=0.3265$ and $\nu=0.6301$\cite{Pelissetto_2002}  (not shown).
In conclusion, we observe a continuous transition with broken U(1)$\times \mathbb{Z}_2$ symmetries 
from a MI phase to  a nematic SF with a director along $z$.

\begin{figure}[t]
	\includegraphics[width=1 \columnwidth]{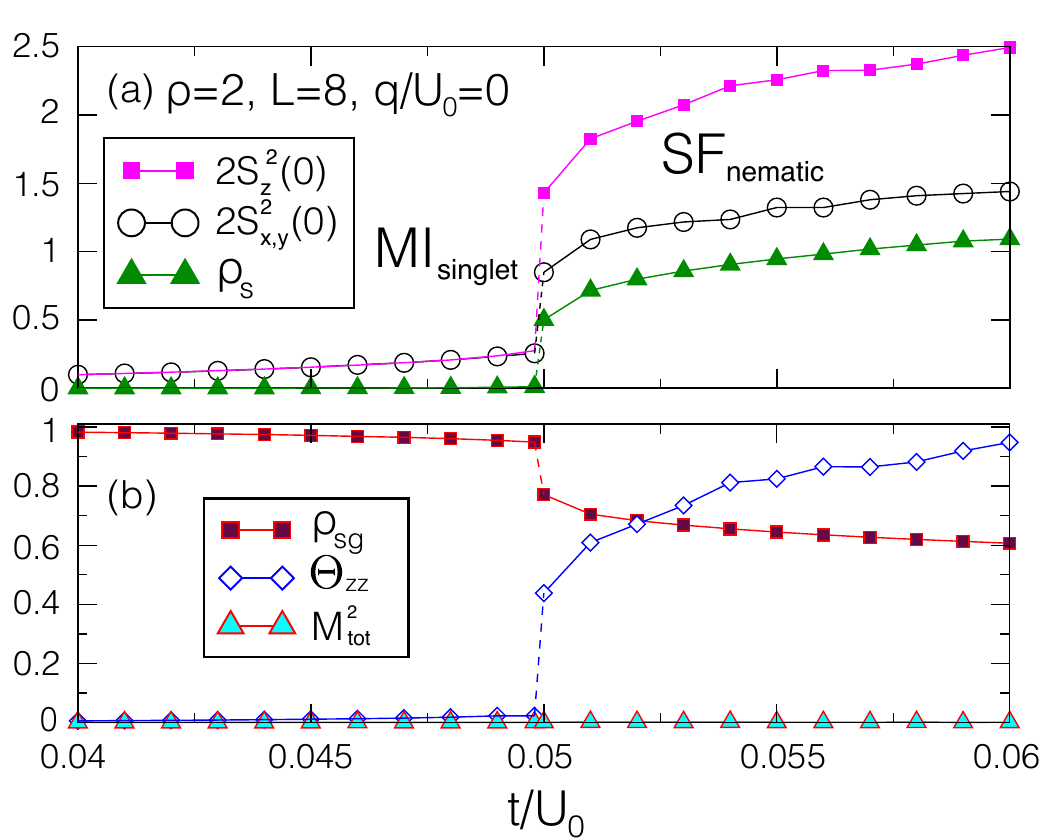}
        \caption {(Color online) QMC data for the $\rho=2$ MI-SF  transition for  $q/U_0=0$ and $L=8$.
        The system undergoes a first-order transition -- indicated by a jump in the quantities -- 
        from a singlet Mott insulator to a fully nematic superfluid.
        }\label{Figure17}
\end{figure}

Finally, we discuss the case $q=0$.
The main difference with the previous cases is that the nematic order could be established along the three axes in the superfluid phase.
In the $t/U_0\to0$ limit, the singlet MI is described by the 
singlet wave function $|\Phi_{\rho=2, q=0}\rangle = \frac{1}{\sqrt{3}} \left(   \sqrt{2} |1,0,1\rangle  -|0,2,0\rangle  \right)$
with  $\rho_{\rm sg}=1$, $\rho_\pm=\rho_0=2/3$, and $S_\alpha^2(0)=0$; see Fig.~\ref{Figure17}.
The jump in $\rho_s$ discussed in Fig.~\ref{Figure14}(b) is also observed in the component of the 
local magnetic moment $S^2_\alpha(0)$, which becomes finite in the three axes $x, y, z$ at the transition at $t_c/U_0\simeq0.05$ [Fig.~\ref{Figure17}(a)].
Furthermore, the singlet density $\rho_{\rm sg}$ and the nematic order parameter $\Theta_{zz}$ also jump at the transition, 
whereas the global magnetization $M^2_{\rm tot}$ remains zero; see Fig.~\ref{Figure17}(b).
Therefore, the nematic order and the phase coherence are simultaneously established when the hopping $t/U_0$ is strong
enough to destroy the singlet state.
Since $M^2_{\rm tot}=0$ and $S^2_\alpha(0)\neq0$ $\forall \alpha$ in the superfluid phase, it is clear that the 
nematic correlations  are finite along the three axes, $M^2_{\alpha}-S_\alpha^2(0) <0$ with $\alpha=\{x, y, z\}$.
In conclusion, the system undergoes a first-order transition from a singlet MI to a fully nematic SF.

\subsection{Vertical slice of the phase diagram}
\label{sec4_subE}

\begin{figure}[t]
	\includegraphics[width=1 \columnwidth]{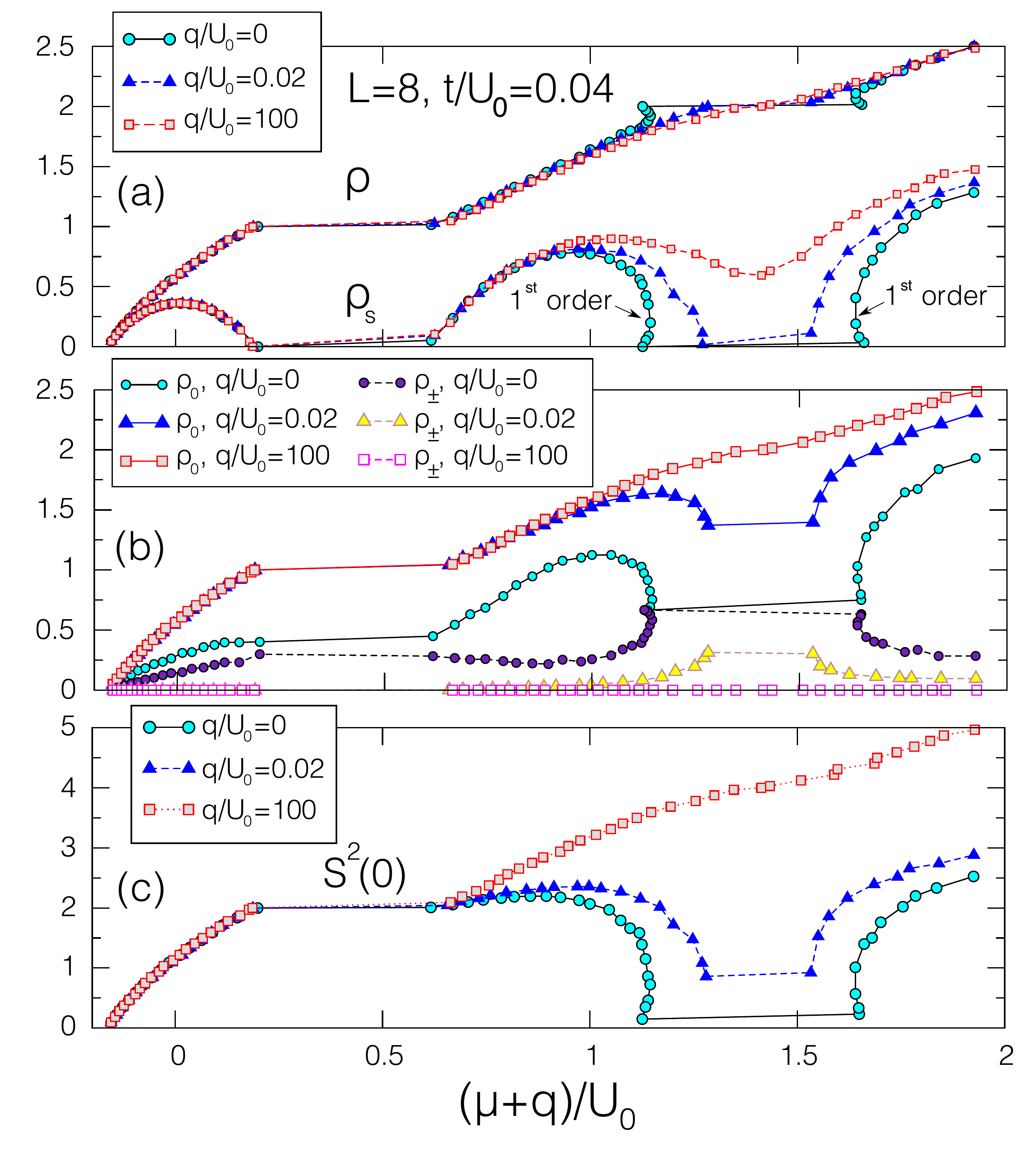}
        \caption {(Color online) QMC data of the vertical slice of phase diagram Fig.~\ref{Figure6}(b) at  fixed hopping  $t/U_0=0.04$
        for $q/U_0=\{0, 0.02, 100\}$. 
        (a) Total density $\rho$, and superfluid density $\rho_s$, (b) densities of each Zeeman state $\rho_\sigma$, and (c) total magnetic local moment $S^2(0)$ as function of $\mu$.}
\label{Figure18}
\end{figure}

To complete the picture, we discuss a vertical slice in the phase diagram Fig.~\ref{Figure6}(b).
Figure~\ref{Figure18} shows such a slice at $t/U_0=0.04$
for three values of $q/U_0$.
For $q=0$, the first and second Mott lobes are indicated by the plateaus $\rho=1, 2$, with a vanishing
superfluid density $\rho_s=0$; see Fig.~\ref{Figure18}(a). 
When turning on the QZE,  the $\rho=2$ Mott gap reduces (e.g., $q/U_0=0.02$) and disappears for  strong enough $q$
(e.g., $q/U_0=100$), thus leaving space to a superfluid phase $\rho_s\neq0$.
This effect is not observed for the $\rho=1$ Mott gap.
The density $\rho_0$ increases with $q$ and saturates at $\rho_0=\rho$ for very large $q$ (e.g., $q/U_0=100$); 
see Fig.~\ref{Figure18}(b).
Nevertheless, for $q/U_0=0.02$, we still observe a mixing in the populations close to the $\rho=2$ Mott plateau due to the minimization 
of $S^2(0)$ which forms singlet state pairs, and therefore populates the  $\sigma=\{\downarrow, \uparrow\}$  states.
Furthermore, the competition between the QZE and the spin-spin interaction is shown in 
Fig.~\ref{Figure18} (c):
for $\rho>1$, the formation of singlet states is activated for $q=0$, thus
leading to the minimization of the local magnetic moment such that $S^2(0)\sim0$ in the $\rho=2$ Mott phase [$S^2(0)$ does not strictly vanish for $t/U_0=0.04$ since the hopping term
destroys a small fraction of singlet pairs]. For $q/U_0=0.02$,  the singlet pairs are partially destroyed by the QZE and  $S^2(0)$ takes a non vanishing value 
in the $\rho=2$ Mott phase. For very large $q$ (e.g., $q/U_0=100$), for which $\rho_0=\rho$,
it is not possible to form singlet state anymore and  the magnetic local moment saturates $S^2(0)=2\rho$; see Eq.~(\ref{magnetic_local_moment_squared}).

Concerning the quantum phase transitions, our QMC simulations predict continuous transitions, except for the $\rho=2$ MI-SF transitions with $q=0$:
the negative slope in $\rho$ versus $\mu$ indicates a negative compressibility since  $\kappa= \partial \rho / \partial \mu < 0$, thus indicating a metastable region. 
This signal, which  is a well-know signature of a first-order transition in the canonical ensemble, \cite{batrouni2000, Deforges_2013, Deforges_2011} 
is also observed in other quantities, e.g., $\rho_\alpha, \rho_s$, and $S^2(0)$.
These QMC results for $q=0$ are in good qualitative agreement with  mean field results of Fig.~\ref{Figure1}.
Nevertheless, these two approaches give incompatible predictions concerning the nature of the $\rho=2$  MI-SF transition for  finite $q$:
According to the mean-field approach, the $\rho=2$ MI-SF transition should be first order for $q/U_0=0.02$.
However, the first-order signature is not observed for $q/U_0=0.02$ with QMC simulation, for which the $\rho=2$ MI-SF transition is continuous; see Fig.~\ref{Figure18}.

\section{Conclusions}
\label{sec_5}  
  
Employing quantum Monte Carlo simulations and mean-field theory, we  derived  the 
phase diagram of interacting lattice spin-1 bosons subject to the QZE. 
The interactions are among the simplest possible for such
a system: an on-site repulsion independent of spin and
an on-site  antiferromagnetic coupling between spins on the same site.
The QZE splits the energy of the sublevels  $\sigma=\pm1$ and $\sigma=0$.

We have particularly focused on the magnetic properties of the Mott and superfluid phases, and on the Mott-superfluid transition when 
varying the Zeeman splitting.
In the absence of QZE, 
the antiferromagnetic  interactions lead to the establishment of a nematic state 
-- i.e., a state breaking  spin-rotation  symmetry  without magnetic order --
in the superfluid phase and in the Mott phases with odd filling, whereas the system adopts a singlet state with zero magnetic local  moment in even Mott lobes.\cite{Deforges_2013} 
Both quantum Monte Carlo simulations and mean-field theory show that the QZE, which directly impacts the populations of $\sigma=\{\pm1, 0\}$ states,  destroys the singlet state at
the tip of the even Mott lobes, thus leaving the space to the superfluid phase. 
This effect is not observed in the Mott lobes with one particle per site since the system cannot form a singlet state.
Therefore, the QZE  acts as a control parameter for the MI-SF transition with even filling and  fixed hopping, 
as observed in a cubic lattice.\cite{Liu_2016} 
Our present study goes beyond the mean-field approximation since  quantum Monte Carlo simulations give
access to magnetic correlation functions required for a reliable definition of a nematic order parameter.
We found a  spin nematic order with director along the $z$ axis in the odd Mott lobes and in the superfluid phase for  favored $\sigma=\pm1$ states, whereas
the $xy$ components of the nematic director remain finite  for moderate QZE in the superfluid phase with even filling.
We also elucidate the nature of the quantum phase transitions:
the Mott-superfluid transition with even filling  is found to be first order for  $q/U_0 \in[-0.005, 0.001]$ and is 3D XY otherwise, contrary
to the mean field approach which predicts a first-order transition in a larger range, for  $q/U_0\leq0.04$.
Our study clearly shows that the QZE is a control parameter for
both the nematic structure and for the MI-SF transition.
This phenomenology sets the stage for future experiments on spinor condensates in
optical lattices, using state-of-the-art techniques.\cite{Zibold_2016, Trotzky_2010, Liu_2016}

\begin{acknowledgments}
We thank Tommaso Roscilde, Fabrice Gerbier, Fr\'ed\'eric Mila,  Christophe Chatelain, Angelika Knothe, and Andreas Buchleitner for useful discussions
and   Fr\'ed\'eric H\'ebert, Tommaso Roscilde, Fabrice Gerbier and  Fr\'ed\'eric Mila for their critical reading of the manuscript. 
The authors acknowledge support by the state of Baden-W\"urttemberg through bwHPC (NEMO and JUSTUS clusters)
and the Alexander von Humboldt-Foundation for financial support.
\end{acknowledgments}

\end{document}